\documentclass[preprint,showpacs,preprintnumbers,amsmath,amssymb]{revtex4}

\usepackage{graphicx}
\usepackage{epstopdf}

\usepackage{dcolumn}
\usepackage{bm}
\usepackage{color}

\usepackage{feynmf}
\usepackage{slashed}
\usepackage{enumerate}
\usepackage{caption}

\usepackage[utf8]{inputenc}

\usepackage[colorlinks=true,urlcolor=blue,anchorcolor=blue,citecolor=blue,filecolor=blue,linkcolor=blue,menucolor=blue,linktocpage=true]{hyperref} 

\begin{document}

\title{Diagrammatic structures of the Nielsen identity}

\author{Yi-Lei Tang}
\thanks{tangylei@mail.sysu.edu.cn}
\affiliation{School of Physics, Sun Yat-Sen University, Guangzhou 510275, China}

\date{\today}

\begin{abstract}
The $\Gamma$-function, or the effective potential of a gauge field theory should comply with the Nielsen identity, which implies how the effective potential evolves as we shift the gauge-fixing term. In this paper, relying on an abelian toy model, we aim at proving this identity in a diagrammatic form with the $\overline{R}_{\xi}$ gauge. The basic idea is to find out the ghost chain after partially differentiating the diagram by the $\xi$ parameter, and shrink the waists of the diagram into points to separate the bulk-part and $C$-part of the diagrams.  The calculations can be generalized to the models implemented with non-abelian groups, multiple Higgs and fermion multiplets, and to the finite temperature cases. Inspired by this, we also suggest that when resumming the super-daisy diagrams, one can deduct some irrelevant terms at the connections between the daisy ringlets to fit the Nielsen identity up to arbitrary $\hbar$ orders. 
\end{abstract}
\pacs{}

\keywords{}

\maketitle
\section{Introduction}

The effective potential is usually utilized to evaluate a set of observables related with the transition between different meta-stable states, or ``vacuums''. Scalar fields are usually regarded as the order parameters, and the effective potential, as the functional of the scalar fields, sometimes induce different local minimums, with their locations and values varying as the temperature evolves. A barrier between two different phases can create a first-order phase transition. During this process, bubbles are created and then expand, with the two vacuums separated by the bubble wall. If this happens in the early universe, the bubble expanding processes might also generate the primary stochastic gravitational waves, and induce the baryon asymmetry as the bubble wall shift through the hot plasma. Within the frameset of the gauge theories and considering the $\xi$-dependent terms, the phase transition rates or temperatures\cite{Metaxas:1995ab, Boyanovsky:1996dc,Garny:2012cg, Endo:2017gal, Chigusa:2018uuj, Endo:2017tsz, Hirvonen:2021zej, Arunasalam:2021zrs, Cho:2021itv, Chiang:2018gsn, Espinosa:2016nld, Lalak:2016zlv, Andreassen:2016cvx,Plascencia:2015pga, Chao:2014ina, Metaxas:2000cw, Baacke:1999sc},  the phase patterns or vacuum stability\cite{Markkanen:2018pdo, DiLuzio:2014bua, Isidori:2001bm}, as well as the primary stochastic gravitational wave relic densities\cite{Chiang:2017zbz, Costa:2022oaa, Chatterjee:2022pxf, Su:2020pjw, Croon:2020cgk, Morais:2019fnm, Chao:2017vrq}, the baryon asymmetry\cite{Patel:2011th, Morrissey:2012db} and the (pole-)mass, mixing parameters or resonance shapes\cite{Shen:2021tqk, Metaxas:2020atd, Maas:2020kda, Kim:2020cms, Dudal:2019pyg, Dudal:2019aew, Siringo:2018uho, Kim:2017jav, Kanemura:2017wtm, Iacobellis:2016eof, Kim:2016rnz, Kim:2015lra, Bian:2013xra, Das:2012qz, Bettinelli:2007eu, Zhou:2005nz, Espinosa:2002cd, Grassi:2001bz, Yamada:2001px, Gambino:1999ai, Gambino:1998ec, Lin:1998up, Breckenridge:1994gs}, the plasma parameters\cite{Kobes:1990dc} are all observables which must be gauge-invariant. However practical gauge independent evaluations are far from the straightforward tasks.

It is well-known that the effective potential (or equivalently, the $\Gamma$-functional) of a gauge field theory model should obey the Nielsen-Fukuda-Kugo identity\cite{Nielsen:1975fs, Fukuda:1975di}(For some alternative discussions, see Ref.~\cite{DelCima:1999dr, Contreras:1996nx, deLima:1989yf, Mitani:1988xm, DoNascimento:1987mn, Johnston:1986ib, Johnston:1984sc, Aitchison:1983ns}). Although the order parameters and the potential energies are generally nonphysical, and might be gauge-dependent quantities, however at the ``extrema'' of the effective potential, such as the minimum of the effective potential of the homogeneous vacuum, or the dynamical bubble solution satisfying the equations of motion, the effective potential values become gauge-independent. Other observables depending on the potential values then become gauge-independent.

The feasible algorithm to evaluate the effective potential is to sum over a particular bunch of one-particle-irreducible (1PI) Feynmann diagrams (or ``amputated'' diagrams), with their external lines connected to the field values selected as order parameters. For example, the widely utilized Coleman-Weinberg potential\cite{Coleman:1973jx} is a result of all one-loop diagrams. Resummation algorithms such as (super-)daisy resummation, renormalization group equation (RGE) improved effective potential might also be applied (See Ref.~\cite{Andreassen:2014eha} for a recent description of the daisy and RGE improved resummation, and the references therein. See Ref.~\cite{Nielsen:2014spa} for an RGE improved example, See Ref.~\cite{Dolan:1973qd, Quiros:1992ez, Curtin:2016urg} for the idea and descriptions of super-daisy diagrams). All these algorithms neglect diagrams. However, the Nielsen identity is derived through the path integral methods, and is a result of a sum of all possible 1PI diagrams.  Therefore it is difficult to acquire a practical effective potential that satisfies the Nielsen identity rigorously. In the literature, usually the effective potential is expanded up to a finite order of loop(s) or $\hbar$\cite{Patel:2011th, Ekstedt:2018ftj, Andreassen:2014eha, Andreassen:2014gha,  Laine:1994zq, Ferreira:2019bij, Alexander:2008hd, Laine:1994bf}, and there is always an unbalance of the $\hbar$ orders on both sides of the equals sign. These remained unbalances sometimes are ascribed to the ``higher orders''.

One might ask the question whether it is possible to acquire an effective potential satisfying the Nielsen identity up to all $\hbar$ orders without evaluating all possible diagrams. To answer this question, it might be beneficial to study the diagrammatic structures of this identity to help us winnow the terms to sum over. In the literature, one-loop or two-loop results have been computed numerically to verify this identity\cite{Metaxas:1995ab, Garny:2012cg, Andreassen:2014gha, Andreassen:2014eha}. In this paper, inspired by the diagrammatic method described in Ref.~\cite{Peskin:1995ev, Ko:2019wxq, JunmouPaper} to prove the Ward identity, we aim at illuminating the diagrammatic structure of the Nielsen identity regardless of the orders or detailed values of the diagrams. We anatomize the diagram structures to see how different seemingly irrelevant terms cancel each other among various related diagrams and the remaining terms exactly satisfy the Nielsen identity up to all orders.  Since we are working on the integrands inside the loop momentum integration, the renormalization and infrared divergence issues\cite{Lewandowski:2013uha, Batalin:2019wkb, Ekstedt:2018ftj, Martin:2018emo, Espinosa:2016uaw, Martin:2014bca} are set aside on this stage., and our results can be easily generalized to finite temperature cases in the frameset of the imaginary time formalism. We believe that these details can help the future researchers verify their results when more precise evaluations will be performed.

We also make a preliminary suggestion to revise the super-daisy resummation algorithm to fit the Nilsen identity. The basic idea is to determine the set of diagrams to sum over (to be called the ``gourd-like'' diagrams), and then drop out the terms that requires the cancellation by other diagrams outside this set (to be called the ``waist-structure breaking'' diagrams). We will just outline the basic idea, and leave the detailed operations and evaluations to our future study.

\section{Basic informations about our  Abelian toy model with a single Higgs boson and two Majorana fermions}

We rely on a gauged $U(1)$ toy model implemented with one vector boson $A^{\mu}$, with its mass endowed by the vacuum expectation value (VEV) of a complex Higgs boson $\Phi$. We also introduce a charged Dirac fermion $\psi$ composed of two Weyl elements $\psi_{L, R}$.  In order to generate a Yukawa coupling, $\Phi$ is assigned with twice the opposite charge of $\psi$.  The Lagrangian is given by
\begin{eqnarray}
\mathcal{L} = -\frac{1}{4} F_{\mu \nu} F^{\mu \nu} + ( D_{\mu} \Phi )^* D^{\mu} \Phi + i \overline{\psi} D\!\!\!\!/ \psi - m_{\psi} \overline{\psi} \psi - V(\Phi, \Phi^{\dagger}) + (\frac{\sqrt{2}}{2} y \Phi \overline{\psi^C} \psi + \text{h.c.}),
\end{eqnarray}
where $F_{\mu \nu} = \partial_{\mu} A_{\nu} - \partial_{\nu} A_{\mu}$, and $D_{\mu} = \partial_{\mu} + i Q_X g A_{\mu}$. $g$ is the gauge coupling constant, while $Q_X$ is the charge that the field $X$ carries. As we have mentioned, we assign $Q_{\Phi} = - 2 Q_{\psi}$ for the validity of the Yukawa $(\frac{\sqrt{2}}{2} y \Phi \overline{\psi^C} \psi + \text{h.c.})$ terms.  For simplicity, we set $Q_{\Phi}=1$ in the rest of our paper.

The general tree-level renormalizable potential $V(\Phi, \Phi^{\dagger})$ is
\begin{eqnarray}
V(\Phi, \Phi^{\dagger}) = \lambda (\Phi^{\dagger} \Phi)^2 + \mu^2 \Phi^{\dagger} \Phi, \label{EffectivePotential}
\end{eqnarray}
where $\lambda$ is the coupling constant. The spontaneously symmetry breaking at the zero temperature requires $\mu^2 < 0$. Decompose the $\Phi$ into its zero-temperature VEV $v$, the real part $R$ and the imaginary part $I$ (also to be called the ``Goldstone''),
\begin{eqnarray}
\Phi = \frac{v + R + i I}{\sqrt{2}},
\end{eqnarray}
so the vector boson becomes massive, and
\begin{eqnarray}
m_A = g v.
\end{eqnarray}
The minimum condition of the (\ref{EffectivePotential}) brings out the mass term of $R$,
\begin{eqnarray}
m_R^2 = 2 \lambda v^2.
\end{eqnarray}
The fermionic $\psi$ is also split into two Majorana components. Decompose $\psi$ into two Weyl spinors
\begin{eqnarray}
\psi = \left[ \begin{array}{c}
\psi_L \\
i \sigma^2 \psi_R^*
\end{array} \right],
\end{eqnarray}
so the mass matrix becomes
\begin{eqnarray}
\mathcal{L} \supset \frac{1}{2} [\psi_L^T \psi_R^T] \left[ \begin{array}{cc}
\delta m & m_{\psi} \\
m_{\psi} & \delta m
\end{array} \right] \left[ \begin{array}{c}
\psi_L \\
\psi_R
\end{array} \right] + \text{h.c.}, \label{PsiMassMatrix}
\end{eqnarray}
where $\delta m = y v$.  (\ref{PsiMassMatrix}) splits the fermions into two mass eigenstates
\begin{eqnarray}
\mathcal{L} \supset \frac{1}{2} [\tilde{\psi}_1^T \tilde{\psi}_2^T] \left[ \begin{array}{cc}
m_{\psi}-\delta m & 0 \\
0 & m_{\psi}+\delta m
\end{array} \right] \left[ \begin{array}{c}
\tilde{\psi}_1 \\
\tilde{\psi}_2
\end{array} \right] + \text{h.c.}, 
\end{eqnarray}
where
\begin{eqnarray}
\tilde{\psi}_1 &=& \frac{i}{\sqrt{2}} ( \psi_L - \psi_R ), \nonumber \\
\tilde{\psi}_2 &=& \frac{1}{\sqrt{2}} ( \psi_L + \psi_R ).
\end{eqnarray}
We also define
\begin{eqnarray}
m_1 &=& m_{\psi}-\delta m, \nonumber \\
m_2 &=& m_{\psi}+\delta m \label{m1m2}
\end{eqnarray}
for future simplicity.
One can then define the 4-component Majorana spinors
\begin{eqnarray}
\psi_i = \left[ \begin{array}{c}
\tilde{\psi}_i \\
i \sigma^2 \tilde{\psi}_i^*
\end{array} \right],~i=1,2.
\end{eqnarray}
in place of the Weyl spinors $\tilde{\psi}_{1,2}$ for more convenient calculations. The Yukawa and gauge interactions are finally rendered into
\begin{eqnarray}
\mathcal{L} \supset \frac{y}{2} ( -\overline{\psi}_1 R \psi_1 + \overline{\psi}_2 R \psi_2 + \overline{\psi}_1 I \psi_2 + \overline{\psi}_2 I \psi_1) + i Q_{\psi} g \overline{\psi}_1 A\!\!\!/ \psi_2.
\end{eqnarray}

In this paper, we adopt the $\overline{R}_{\xi}$ gauge\cite{Kastening:1993zn} where the VEV $v$ always adjoin with $R$ in the gauge fixing terms
\begin{eqnarray}
\mathcal{L}_{\text{g.f.}} &=& -\frac{1}{2 \xi} F^2, \label{GaugeFixing} \\
F &=& \partial_{\mu} A^{\mu} - g \frac{ \Phi^2 - (\Phi^{\dagger})^2}{2 i}= \partial_{\mu} A^{\mu} - g \xi (v+R) I,
\end{eqnarray}
inducing the Faddeev-Popov ghost  interactions
\begin{eqnarray}
\mathcal{L}_{\text{f.p.}} = - \overline{c} [ \square + \xi g^2 (v+R)^2 - \xi g^2 I^2  ] c. \label{GhostInteraction}
\end{eqnarray}
Notice that (\ref{GaugeFixing}) shifts the  $R$-$R$-$I$-$I$, $R$-$R$-$I$, and $R$-$I$-$A$ couplings, and (\ref{GhostInteraction}) leads to quite different ghost interaction terms compared with the more familiar $R_{\xi}$ gauges, in which the VEV $v$ is hard-coded into the gauge fixing terms. The $\overline{R}_{\xi}$ gauges will help us avoid the intricate mixings between the longitudinal vector bosons and the Goldstone bosons when calculating the effective potential expanded from any point in the field space, with the price of the more complicated interaction terms.  In this paper, we expand from the minimum of the effective potential defined in (\ref{EffectivePotential}), and we should note that most of our derivations in this paper can be easily transplanted and reformulated to the $R_{\xi}$ gauges, so our proof is valid in both these gauges.

We now enumerate the Feynmann rules of this model. The propagators are
\begin{eqnarray}
\vcenter{\hbox{\includegraphics[width=2in]{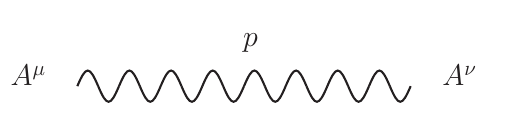}}} &=& \frac{-i}{p^2-m_A^2} \left[ g^{\mu \nu} - \frac{p^{\mu} p^{\nu}}{p^2-\xi m_A^2} (1-\xi) \right], \\
\vcenter{\hbox{\includegraphics[width=2in]{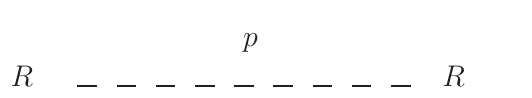}}} &=& \frac{i}{p^2-m_A^2}, \\
\vcenter{\hbox{\includegraphics[width=2in]{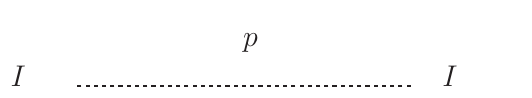}}} &=& \frac{i}{p^2-\xi m_A^2},  \\
\vcenter{\hbox{\includegraphics[width=2in]{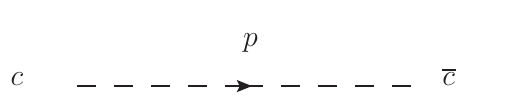}}} &=& \frac{i}{p^2-\xi m_A^2},  \\
\vcenter{\hbox{\includegraphics[width=2in]{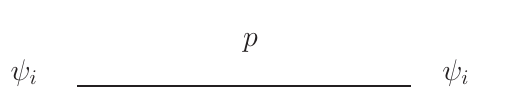}}} &=& \frac{i (p\!\!\!/+m_i)}{p^2-\xi m_i^2}~~(i=1,2),  
\end{eqnarray}
The gauge vertices involving scalars and ghosts are given by
\begin{eqnarray}
\vcenter{\hbox{\includegraphics[width=2in]{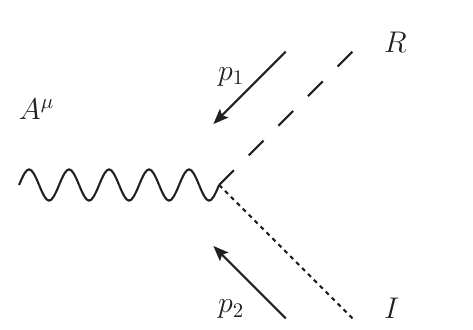}}} &=& 2 g p_1^{\mu},  \label{V_ARI}\\
\vcenter{\hbox{\includegraphics[width=2in]{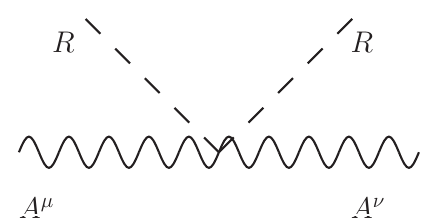}}} \vcenter{\hbox{\includegraphics[width=2in]{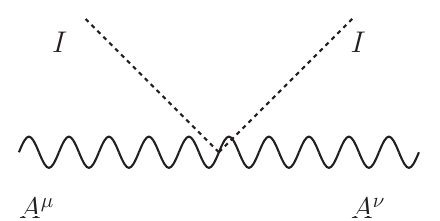}}} &=& 2 g^2 i g^{\mu \nu},  \label{AASS}\\
\vcenter{\hbox{\includegraphics[width=2in]{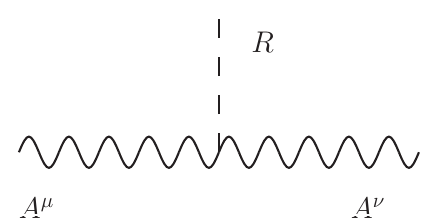}}} &=& 2 i g^2 v g^{\mu \nu}, \\
\vcenter{\hbox{\includegraphics[width=2in]{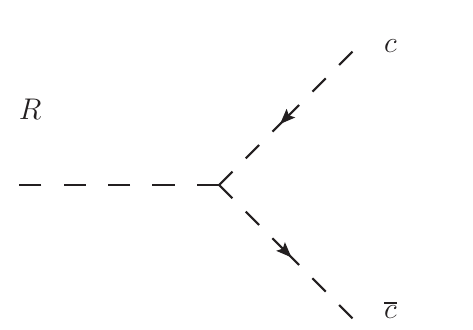}}} &=& -2 i g^2 v \xi, \label{V_Rcc} \\
\vcenter{\hbox{\includegraphics[width=2in]{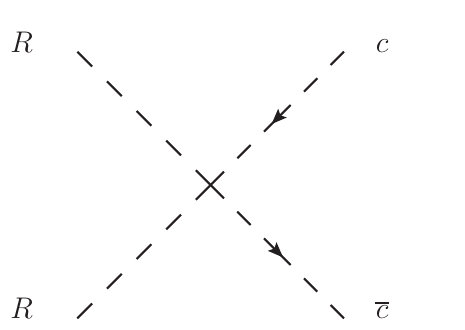}}} &=& -2 i g^2 \xi, \label{V_RRcc} \\
\vcenter{\hbox{\includegraphics[width=2in]{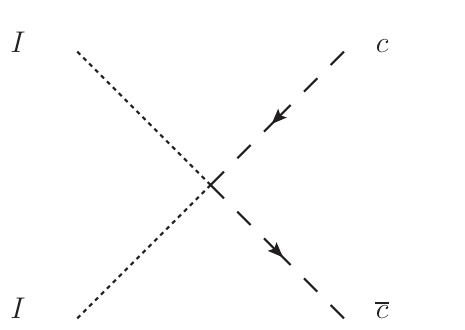}}} &=& 2 i g^2 \xi, \label{V_IIcc} 
\end{eqnarray}
The scalars self-interact through these vertices,
\begin{eqnarray}
\vcenter{\hbox{\includegraphics[width=2in]{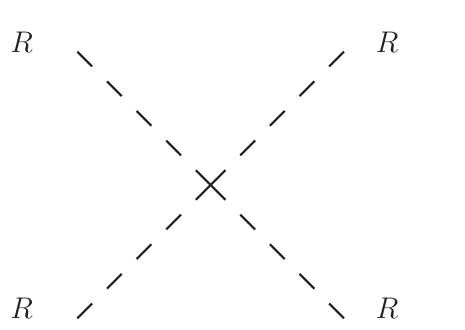}}}  \vcenter{\hbox{\includegraphics[width=2in]{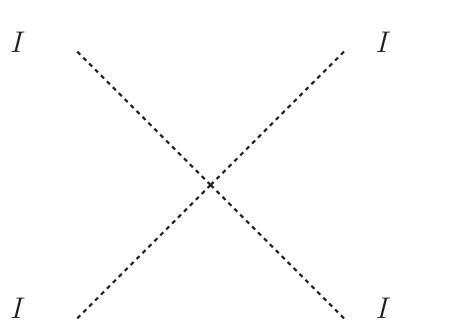}}} &=& -6 i \lambda,  \label{RRRR_IIII_Vertex}\\
\vcenter{\hbox{\includegraphics[width=2in]{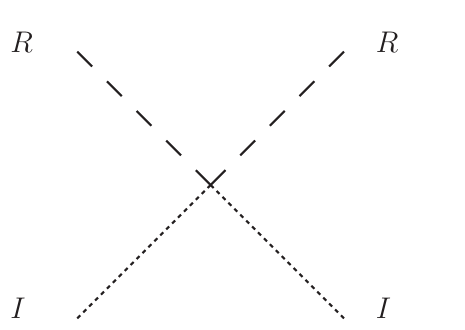}}} &=& -2 i (\lambda + \xi g^2),  \label{RRII_Vertex}\\
\vcenter{\hbox{\includegraphics[width=2in]{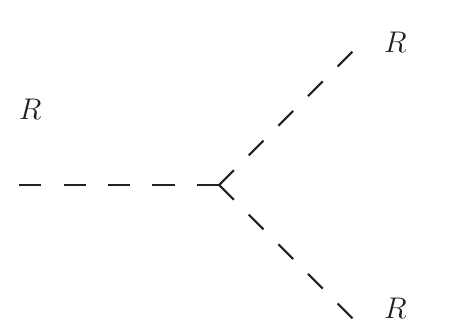}}} &=& -6 i \lambda v, \\
\vcenter{\hbox{\includegraphics[width=2in]{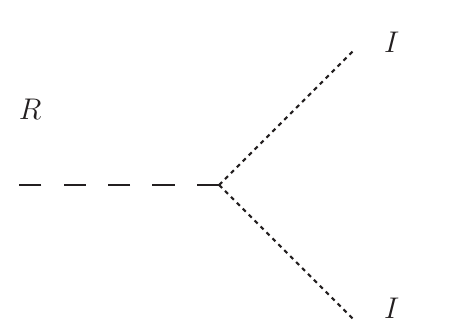}}} &=& -2 i (\lambda + \xi g^2) v. \label{RII_Vertex}
\end{eqnarray}
Finally, the fermions are involved in the following vertices,
\begin{eqnarray}
\vcenter{\hbox{\includegraphics[width=2in]{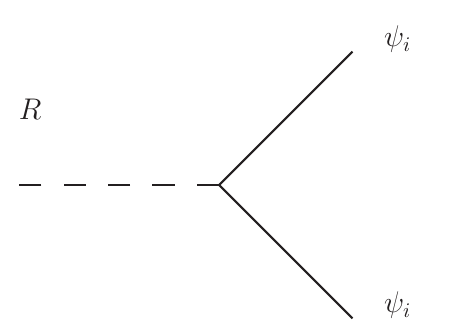}}} &=& (-\delta_{i 1}+\delta_{i 2}) y i, \\
\vcenter{\hbox{\includegraphics[width=2in]{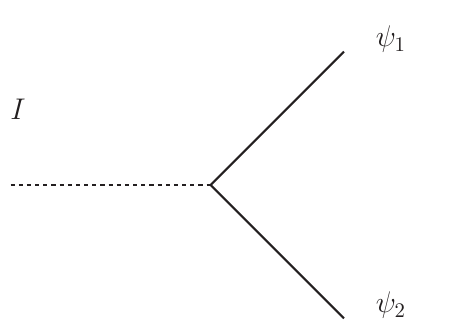}}} &=&  y i,  \\
\vcenter{\hbox{\includegraphics[width=2in]{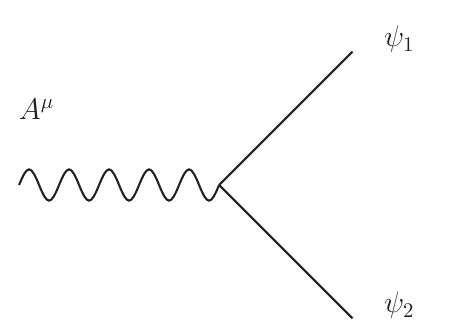}}} &=&  -Q_{\psi} g \gamma^{\mu}, 
\end{eqnarray}
where when the symbol $i$ acts as an index, it refers to $i=1,2$, and otherwise it symbolizes the imaginary unit.

\section{Diagrammatic proof of the Nielsen identity}
\subsection{Overview of the Nielsen identity and its diagrammatic counterpart}

The functional $\Gamma$, which is the summation of all the 1PI diagrams, complies the Nielsen identity if one changes the gauge fixing condition\cite{Garny:2012cg},
\begin{eqnarray}
\delta \Gamma[\phi] = i \int d^4 x \int d^4 y \frac{\delta \Gamma}{\delta \phi_i(i)} \langle \delta_g \phi_i (x) c(x) \overline{c}(y) \delta^{\prime} F(y) \rangle_{\text{1PI}}, \label{NielsenOrigin}
\end{eqnarray}
where $\phi_i$ runs over all the fields, $\delta_g$ is the generator operator of the gauge field, and $\delta^{\prime} F = \delta F - \frac{F}{2 \xi} \delta \xi$ including both the contributions from shifting $F$ or $\xi$. In this paper, we have for the scalars,
\begin{eqnarray}
\delta_g R = - g I.
\end{eqnarray}
Usually, The physical observables, e.g., tunneling rates, gravitational wave relics, etc.,  that people calculate are at the background configuration that all the VEVs of the vectors, spinors, Goldstones and ghost fields vanish, so only $R$ appears to replace $\phi_i$ at the right-hand side of (\ref{NielsenOrigin}). If we only consider a change in the $\xi$, (\ref{NielsenOrigin}) is reduced to
\begin{eqnarray}
\xi \frac{\partial \Gamma[R, \xi]}{\partial \xi} = -\int d^4 x \frac{\delta \Gamma}{\delta R(x)} C_R(x), \label{NielsenRelying}
\end{eqnarray}
where
\begin{eqnarray}
C_R(x) &=& -\frac{i}{2} \int d^4 y \langle I(x) c(x) \overline{c}(y) (F(y) - 2 \xi \frac{\partial F(y)}{\partial \xi}) \rangle_{\text{1PI}} \nonumber \\
&=& -\frac{i}{2} \int d^4 y \langle I(x) c(x) \overline{c}(y) (\partial_{\mu} A^{\mu} + g \xi (v+R) I) \rangle_{\text{1PI}}. \label{CPartOfNielsen}
\end{eqnarray}
This is the formalism of the Nielsen identity that we are going to verify diagrammatically in this paper.

The partially differentiating operations by $\xi$ at the left-hand side of (\ref{NielsenRelying}) finally exert on the $\xi$-depending propagators and vertices. The vector, Goldstone and ghost propagators differentiated by $\xi$ become
\begin{eqnarray}
\frac{\partial}{\partial \xi} \vcenter{\hbox{\includegraphics[width=2in]{PropagatorA-eps-converted-to.pdf}}} &=&  \vcenter{\hbox{\includegraphics[width=2in]{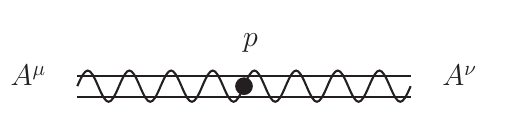}}}  = \frac{-i p^{\mu} p^{\nu}}{(p^2- \xi m_A^2)^2},  \label{dAdXi} \\
\frac{\partial}{\partial \xi} \vcenter{\hbox{\includegraphics[width=2in]{PropagatorGhost-eps-converted-to.pdf}}} &=&  \vcenter{\hbox{\includegraphics[width=2in]{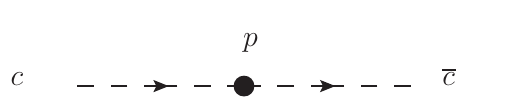}}}  = \frac{i m_A^2}{(p^2- \xi m_A^2)^2},  \label{dPropgatorGhost}\\
\frac{\partial}{\partial \xi} \vcenter{\hbox{\includegraphics[width=2in]{PropagatorI-eps-converted-to.pdf}}} &=&  \vcenter{\hbox{\includegraphics[width=2in]{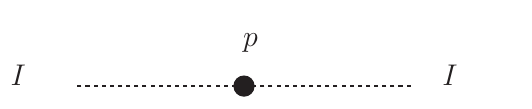}}}  = \frac{i m_A^2}{(p^2- \xi m_A^2)^2}. \label{dPropgatorI}
\end{eqnarray}
The derivative of the corresponding vertices become
\begin{eqnarray}
\frac{\partial}{\partial \xi} \vcenter{\hbox{\includegraphics[width=2in]{V_RRII-eps-converted-to.pdf}}} &=&  \vcenter{\hbox{\includegraphics[width=2in]{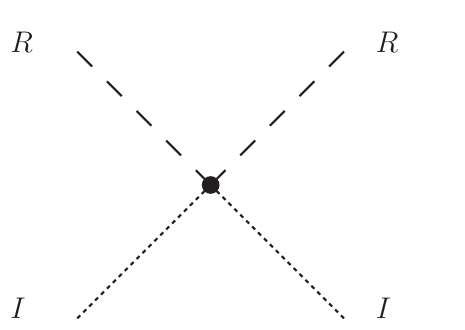}}}  = -2 i g^2,  \label{D_V_RIII} \\
\frac{\partial}{\partial \xi} \vcenter{\hbox{\includegraphics[width=2in]{V_RII-eps-converted-to.pdf}}} &=&  \vcenter{\hbox{\includegraphics[width=2in]{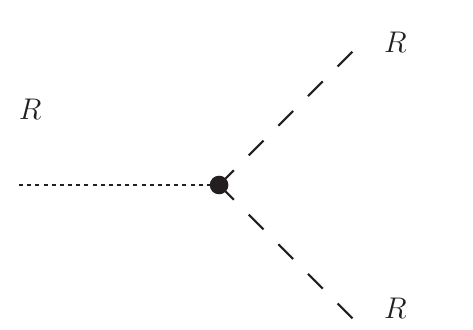}}}  = -2 i g^2 v,  \label{D_V_RII} \\
\frac{\partial}{\partial \xi} \vcenter{\hbox{\includegraphics[width=2in]{V_Rcc-eps-converted-to.pdf}}} &=&  \vcenter{\hbox{\includegraphics[width=2in]{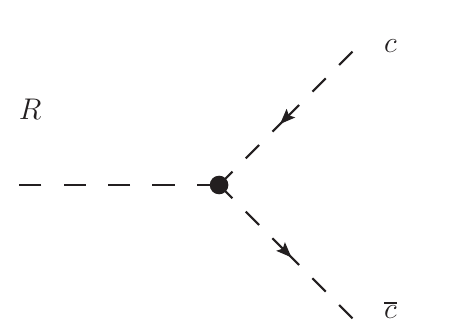}}}  = -2 i g^2 v, \label{D_V_Rcc} \\
\frac{\partial}{\partial \xi} \vcenter{\hbox{\includegraphics[width=2in]{V_RRcc-eps-converted-to.pdf}}} &=&  \vcenter{\hbox{\includegraphics[width=2in]{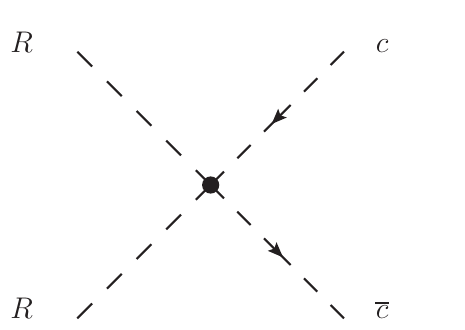}}}  = -2 i g^2. \label{D_V_RRcc} \\
\frac{\partial}{\partial \xi} \vcenter{\hbox{\includegraphics[width=2in]{V_IIcc-eps-converted-to.pdf}}} &=&  \vcenter{\hbox{\includegraphics[width=2in]{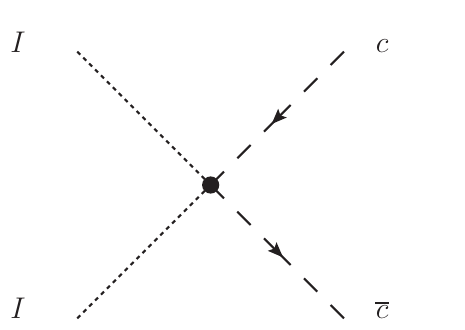}}}  = 2 i g^2. \label{D_V_IIcc}
\end{eqnarray}

\begin{figure}
\begin{equation}
\vcenter{\hbox{\includegraphics[width=3in]{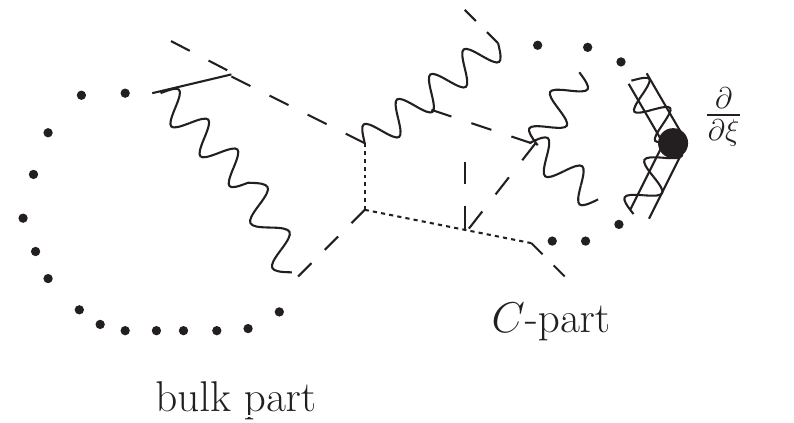}}} {\Huge \Rightarrow}
\vcenter{\hbox{\includegraphics[width=3in]{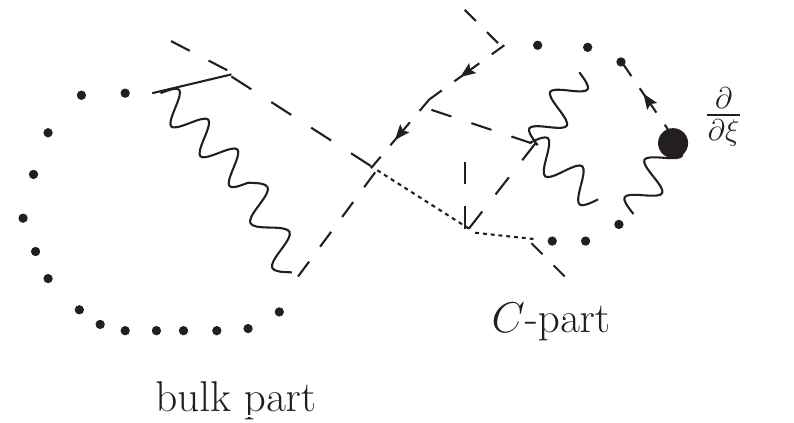}}} \nonumber
\end{equation}
\caption{The sketched process of our proof. }\label{GourdSketch}
\end{figure}

Observe the right-hand side of (\ref{NielsenRelying}). The diagrams composing it look like a gourd with the $\frac{\delta \Gamma}{\delta R}$ part and the $C_R(x)$ part,  which we call the ``bulk part'' and the ``$C$-part'' respectively, connecting through a point-like vertex at the ``waists''.  In the following part of this paper, we will see that these gourd structures are exactly inherited from a group of gourd-shaped diagrams composing the left-hand side of (\ref{NielsenRelying}), with the bulk and the $C$-part sharing only one common vertex or one common internal line as the waists.  In this paper, we define that a ``waist'' should only include one common internal line or vertex. As we show in Fig.~\ref{GourdSketch}, the left panel is one of the diagrams from the left-hand side of the (\ref{NielsenRelying}), with one of the vector propagator differentiated by $\xi$. We will see that this differentiated vector propagator transmutes into a half-ghost propagator,  inducing a ghost-chain arriving at the waist of the gourd to shrink the common internal line shared by the bulk part and $C$-part into a point. This conforms exactly the diagrams implied by the right-hand side of the (\ref{NielsenRelying}).  We will prove the mutual correspondence of the diagrams at both sides of the (\ref{NielsenRelying}) through this process, to illustrate the perturbative structure of the Nielsen identity from a diagrammatic aspect.

\subsection{Isolating the $C$-part of the diagrams}

Let us start from the derivative of the vector propagator (\ref{dAdXi}). If its right part emits an $R$,
\begin{eqnarray}
\vcenter{\hbox{\includegraphics[width=2in]{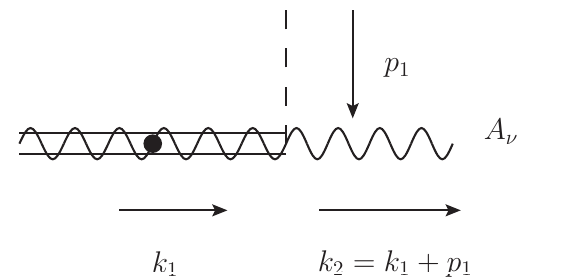}}} = 2 i g^2 v k_1^{\mu}  \frac{-i}{k_2^2-m_A^2} \left[ g_{\mu \nu}-\frac{k_{2 \mu} k_{2 \nu}}{k_2^2 - \xi m_A^2} (1-\xi) \right]. \label{FirstStepR}
\end{eqnarray}
Here we only preserve the $k_1^{\mu}$ factor of the derivative of the vector propagator for abbreviation. Decompose $k_1^{\mu} = k_2^{\mu} - p_1^{\mu}$, we have
\begin{eqnarray}
& & 2 i g^2 v (k_2^{\mu} - p_1^{\mu})  \frac{-i}{k_2^2-m_A^2} \left[ g_{\mu \nu}-\frac{k_{2 \mu} k_{2 \nu}}{k_2^2 - \xi m_A^2} (1-\xi) \right] \nonumber \\
&=& (-2 i g^2 v \xi) \frac{i}{k_2^2 - \xi m_A^2} k_{2 \nu} - i m_A \cdot 2 g p_1^{\mu}  \frac{-i}{k_2^2-m_A^2} \left[ g_{\mu \nu}-\frac{k_{2 \mu} k_{2 \nu}}{k_2^2 - \xi m_A^2} (1-\xi) \right]. \nonumber \\
&=& \vcenter{\hbox{\includegraphics[width=2in]{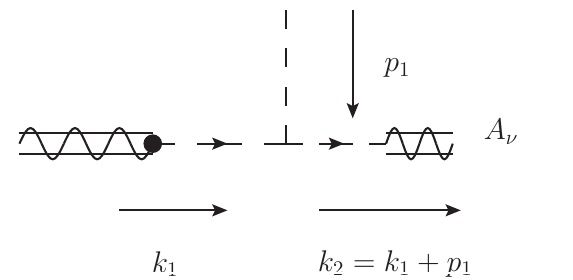}}} - i m_A \vcenter{\hbox{\includegraphics[width=2in]{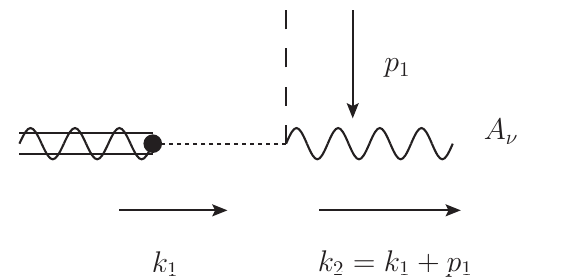}}} . \label{FirstStepR_Decompose}
\end{eqnarray}
The first term looks like a ghost coupling, however the propagator $\frac{i}{k_2^2 - \xi m_A^2}$ is accompanied with an extra $k_{2 \nu}$, which will contract with the index of the following propagators through the coupling in which it still acts as a vector boson. Such a half-ghost half-vector propagator is denoted by
\begin{eqnarray}
\vcenter{\hbox{\includegraphics[width=1.5in]{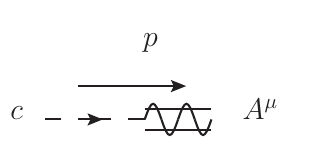}}} = \frac{i p^{\mu}}{p^2-\xi m_A^2}. \label{HalfCHalfA}
\end{eqnarray}
Conveniently, one can decompose this propagator into two parts, the ghost-half $\frac{i}{p^2-\xi m_A^2}$, as well as the vector-half $p^{\mu}$. Therefore we formally define
\begin{eqnarray}
\vcenter{\hbox{\includegraphics[width=0.9in]{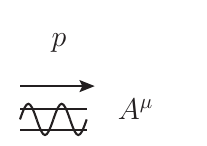}}} = p^{\mu}, \label{HalfA}
\end{eqnarray}
so
\begin{eqnarray}
\vcenter{\hbox{\includegraphics[width=1.5in]{PropagatorHalfCHalfA-eps-converted-to.pdf}}} = \vcenter{\hbox{\includegraphics[width=0.8in]{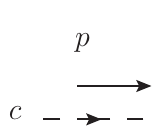}}} \times
\vcenter{\hbox{\includegraphics[width=0.9in]{HalfA-eps-converted-to.pdf}}} = \frac{i}{p^2-\xi m_A^2} \times p^{\mu}. 
\end{eqnarray}
With these convention, we can also omit the left half of (\ref{dAdXi}) during the calculation to reinterpret the $k_1^{\mu}$ appeared in (\ref{FirstStepR}). 

If two $R$'s are emitted,  the following diagrams should be summed over
\begin{eqnarray}
& & \vcenter{\hbox{\includegraphics[width=2.5in]{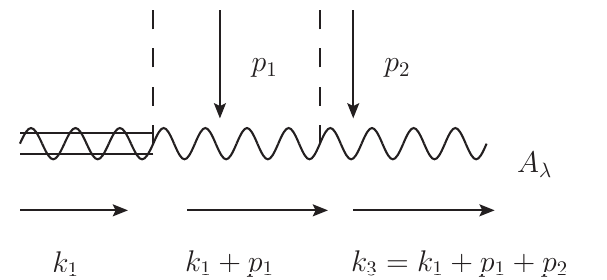}}} + \vcenter{\hbox{\includegraphics[width=2.5in]{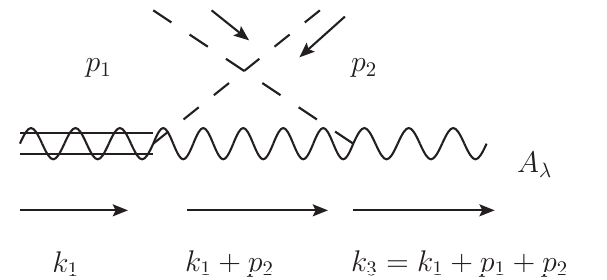}}} \label{FirstStepRR_A}\\
&+& \vcenter{\hbox{\includegraphics[width=2.5in]{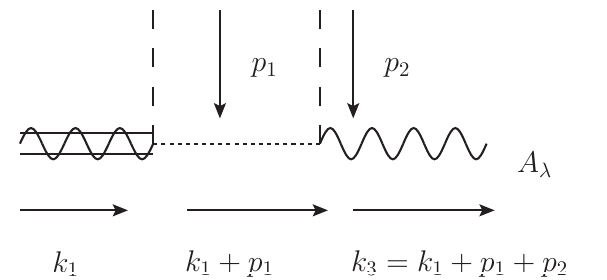}}} + \vcenter{\hbox{\includegraphics[width=2.5in]{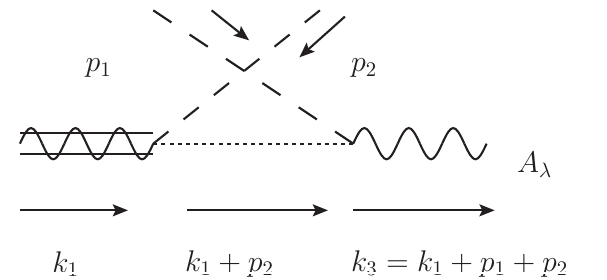}}} \label{FirstStepRR_IMediate}\\
&+&  \vcenter{\hbox{\includegraphics[width=1.9in]{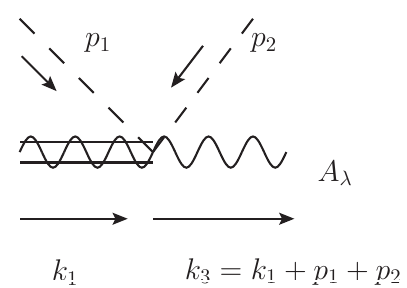}}}. \label{FirstStepRR_4Vertex}
\end{eqnarray}
Directly evaluating these diagrams is cumbersome. Notice that the two diagrams in (\ref{FirstStepRR_A}) can be treated as a direct successor of (\ref{FirstStepR}) by adhering another tail after it, so we utilize (\ref{FirstStepR_Decompose}) to decompose (\ref{FirstStepRR_A}) into two parts
\begin{eqnarray}
& & \vcenter{\hbox{\includegraphics[width=2.5in]{AFirstStepRR-eps-converted-to.pdf}}} + \vcenter{\hbox{\includegraphics[width=2.5in]{AFirstStepRR_AT-eps-converted-to.pdf}}}  \nonumber \\
&=& \vcenter{\hbox{\includegraphics[width=2.5in]{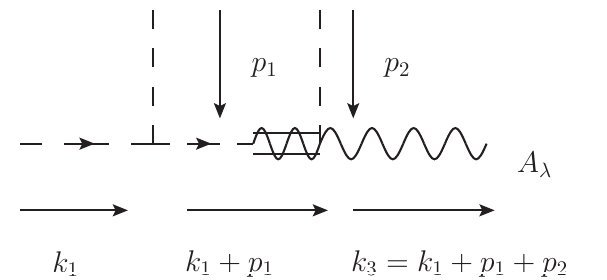}}} + \vcenter{\hbox{\includegraphics[width=2.5in]{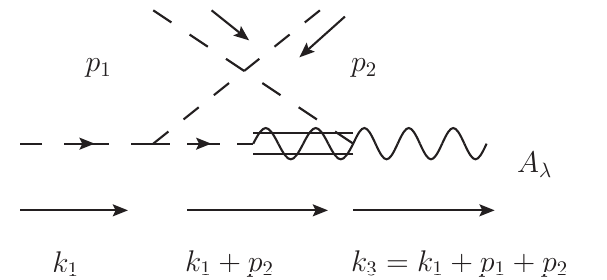}}} \label{GhostToExtend}\\
&+& \left( \vcenter{\hbox{\includegraphics[width=2.5in]{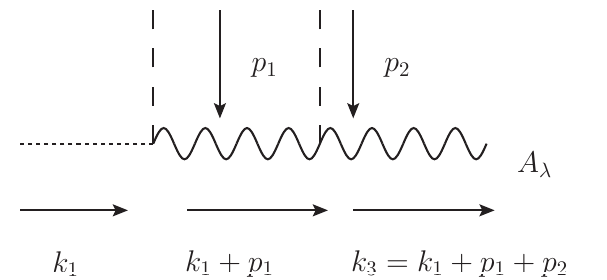}}} + \vcenter{\hbox{\includegraphics[width=2.5in]{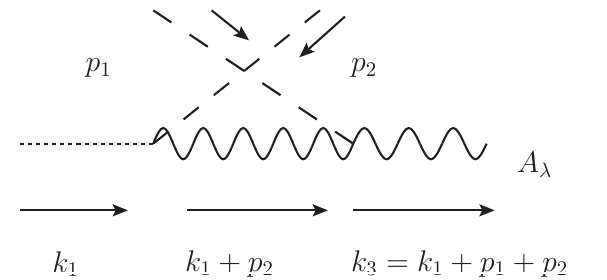}}} \right) \label{IStartPart_RR} \\
&\cdot& (-i m_A) \nonumber
\end{eqnarray}
We now consider the (\ref{GhostToExtend}) terms. Similarly to the processes in (\ref{FirstStepR}-\ref{FirstStepR_Decompose}), the half-vector part of the middle propagator can be further decomposed into two parts. One recursively renders the $k_3$ propagator into a half-ghost half-vector propagator with itself transmuting into a complete ghost propagator. The other part looks like a half-ghost half-$I$ propagator. With the diagrammatic language, it is
\begin{eqnarray}
& & \vcenter{\hbox{\includegraphics[width=2.5in]{AFirstStepRR_cPart-eps-converted-to.pdf}}} + \vcenter{\hbox{\includegraphics[width=2.5in]{AFirstStepRR_AT_cPart-eps-converted-to.pdf}}}  \nonumber \\
&=& \vcenter{\hbox{\includegraphics[width=2.5in]{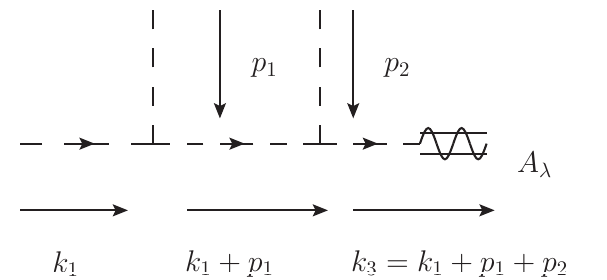}}} + \vcenter{\hbox{\includegraphics[width=2.5in]{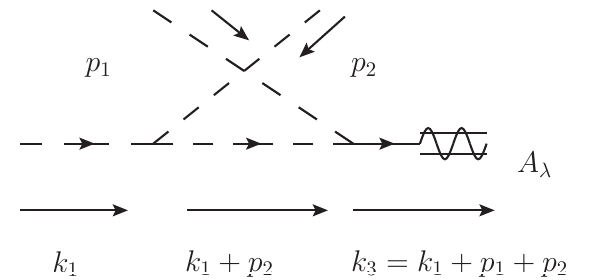}}} \\
&+& \vcenter{\hbox{\includegraphics[width=2.5in]{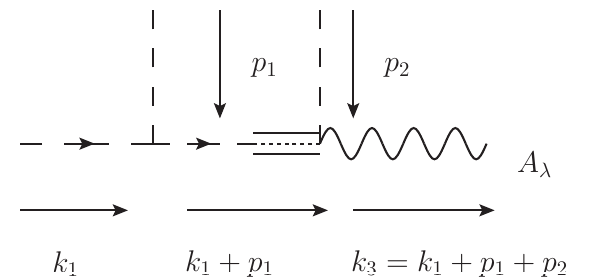}}} + \vcenter{\hbox{\includegraphics[width=2.5in]{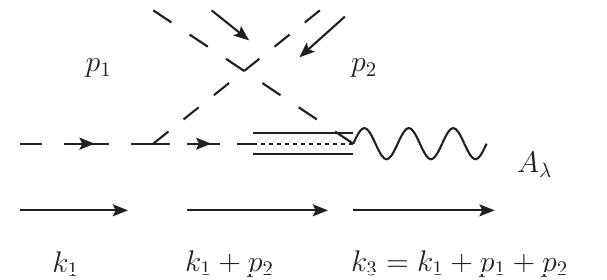}}} \label{FirstStepRR_HalfCHalfI}
\end{eqnarray}
Here the half-ghost half-$I$ propagator is define as
\begin{eqnarray}
\vcenter{\hbox{\includegraphics[width=1.5in]{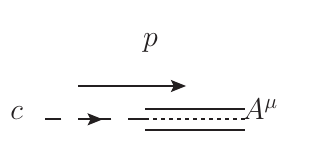}}} = \frac{i(-i m_A)}{p^2-\xi m_A^2}, \label{HalfCHalfIPropagator}
\end{eqnarray}
and the left vertex participates the ghost interactions,  the right vertex participates the $I$ interactions.
Again,
\begin{eqnarray}
\vcenter{\hbox{\includegraphics[width=0.9in]{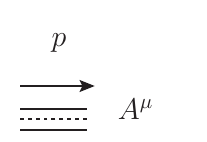}}} = -i m_A. \label{HalfIPropagator}
\end{eqnarray}

Now we calculate the diagrams in (\ref{FirstStepRR_IMediate}), (\ref{FirstStepRR_4Vertex}) and (\ref{FirstStepRR_HalfCHalfI}). Neglect the common $\frac{-i}{k_3^2 - m_A^2} \left[g_{\mu \lambda} - \frac{k_{3 \mu} k_{3 \lambda}}{k_3^2-\xi m_A^2} (1-\xi) \right]$ propagator term for abbreviation, They are
\begin{eqnarray}
\vcenter{\hbox{\includegraphics[width=2.5in]{AFirstStepRR_ccIPart-eps-converted-to.pdf}}} &=& -2i g^2 v \xi \frac{i}{(k_1+p_1)^2 - \xi m_A^2} (-2 i g^2 v p_2^{\mu}) \label{FirstStepRR_Cs} \\ 
\vcenter{\hbox{\includegraphics[width=2.5in]{AFirstStepRR_AT_ccIPart-eps-converted-to.pdf}}} &=& -2i g^2 v \xi \frac{i}{(k_1+p_2)^2 - \xi m_A^2} (-2 i g^2 v p_1^{\mu}) \label{FirstStepRR_Ct} \\
\vcenter{\hbox{\includegraphics[width=2.5in]{AFirstStepRR_I-eps-converted-to.pdf}}} &=& 2 g (p_1 \cdot k_1) \frac{i}{(k_1+p_1)^2 - \xi m_A^2} (2 g p_2^{\mu}) \label{FirstStepRR_Is} \\
\vcenter{\hbox{\includegraphics[width=2.5in]{AFirstStepRR_IT-eps-converted-to.pdf}}} &=& 2 g (p_2 \cdot k_1) \frac{i}{(k_1+p_2)^2 - \xi m_A^2} (2 g p_1^{\mu}) \label{FirstStepRR_It} \\
\vcenter{\hbox{\includegraphics[width=1.9in]{AFirstStepRR_4Vertex-eps-converted-to.pdf}}} &=& 2 g^2 i k_1^{\mu}. \label{FirstStepRR_4VertexResult} 
\end{eqnarray}
Notice that (\ref{FirstStepRR_Ct}) and (\ref{FirstStepRR_Is}) share the same structure with (\ref{FirstStepRR_Cs}) and (\ref{FirstStepRR_It}) except the exchange of $p_1$ and $p_2$,  so we focus on (\ref{FirstStepRR_Cs}) + (\ref{FirstStepRR_Is}), which gives the result
\begin{eqnarray}
& & g [2 (p_1 \cdot k_1) - 2 \xi m_A^2] \frac{i}{(k_1+p_1)^2 - \xi m_A^2} (2 g p_2^{\mu}) \nonumber \\
&=& g [(p_1+k_1)^2-p_1^2-k_1^2- 2 \xi m_A^2] \frac{i}{(k_1+p_1)^2 - \xi m_A^2} (2 g p_2^{\mu}) \nonumber \\
&=& 2 i g^2 p_2^{\mu} - g (p_1^2 - m_R^2) \frac{i}{(k_1+p_1)^2 - \xi m_A^2} (2 g p_2^{\mu}) \nonumber \\
& & - g (k_1^2 - \xi m_A^2)  \frac{i}{(k_1+p_1)^2 - \xi m_A^2} (2 g p_2^{\mu}) - \frac{i g(m_R^2  + 2 \xi m_A^2)}{(k_1+p_1)^2 - \xi m_A^2 } (2 g p_2^{\mu}). \label{FirstStepRR_ToBeCancelled}
\end{eqnarray}
If the $p_1$ propagator is also an internal line, the second term will kill the $p_1^2=m_R^2$ pole of it. Similarly to the diagrammatic proof of the Ward-Takahashi identity in a general phase\cite{Peskin:1995ev, Ko:2019wxq, JunmouPaper},  there must exist another diagram canceling this term.  We will also encounter an example in the following text. On the other hand, when $p_1$ is an external line connecting to the VEV function, we will see that it is the tree level terms $(-\partial^2 - m_R^2) R \in \frac{\partial \Gamma}{\partial R}$ adhering to the loop-level $C$-part components. We also leave this to our future discussion. The $(k_1^2 - \xi m_A^2)$ factor in the third term of (\ref{FirstStepRR_ToBeCancelled}) cancels the previous $k_1^2 = \xi m_A^2$ pole in the $k_1$ propagator, and will finally contribute to the $\langle g \xi RI \rangle_{\text{1PI}}$ term in the (\ref{CPartOfNielsen}). Such kind of cancellation processes is a useful trick, and will reproduce again and again in this paper. For the first term of (\ref{FirstStepRR_ToBeCancelled}),  (\ref{FirstStepRR_Ct})+(\ref{FirstStepRR_It}) will also contribute a $2 i g^2 p_1^{\mu}$. Combined these with the (\ref{FirstStepRR_4VertexResult}), and supplement the last $k_3$ propagator, one obtains
\begin{eqnarray}
& & 2 g^2 i k_3^{\mu} \frac{-i}{k_3^2 - m_A^2} \left[g_{\mu \lambda} - \frac{k_{3 \mu} k_{3 \lambda}}{k_3^2-\xi m_A^2} (1-\xi) \right] = (-2 i g^2  \xi) \frac{i}{k_3^2 - \xi m_A^2} k_{3 \lambda}.  \\
&=& \vcenter{\hbox{\includegraphics[width=1.9in]{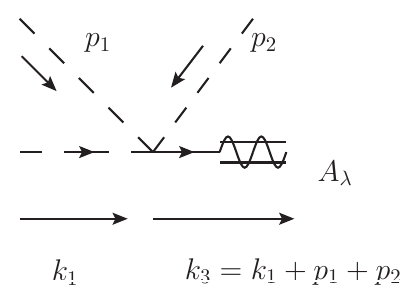}}}. \nonumber
\end{eqnarray}
The last term of (\ref{FirstStepRR_ToBeCancelled}) picks up the factor $g(m_R^2 + 2 \xi m_A^2) = m_A ( 2 \lambda v + 2 g^2 v \xi)$, which is in accordance with the vertex (\ref{RII_Vertex}).  So the corresponding diagram can be depicted as
\begin{eqnarray}
& & -i m_A \frac{2 \lambda v + 2 g^2 v \xi}{(k_1+p_1)^2-\xi m_A^2} (2 g p_2^{\mu}) \frac{-i}{k_3^2 - m_A^2} \left[g_{\mu \lambda} - \frac{k_{3 \mu} k_{3 \lambda}}{k_3^2-\xi m_A^2} (1-\xi) \right] \\
&=& -i m_A \left( \vcenter{\hbox{\includegraphics[width=2.5in]{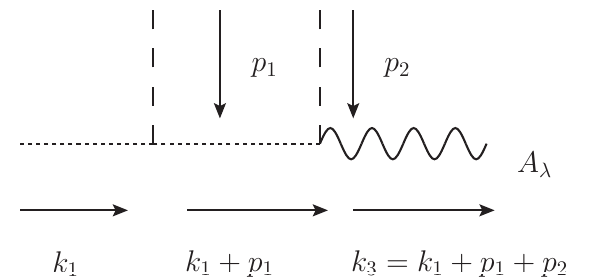}}} \right). \label{FirstStepRR_IIV} \nonumber
\end{eqnarray}

Therefore, finally (\ref{FirstStepRR_A})+(\ref{FirstStepRR_IMediate})+(\ref{FirstStepRR_4Vertex}) can be reduced to
\begin{eqnarray}
& & (\ref{FirstStepRR_A})+(\ref{FirstStepRR_IMediate})+(\ref{FirstStepRR_4Vertex}) \nonumber \\
& \doteq &  \vcenter{\hbox{\includegraphics[width=2.5in]{AFirstStepRR_ccPart-eps-converted-to.pdf}}} + \vcenter{\hbox{\includegraphics[width=2.5in]{AFirstStepRR_AT_ccPart-eps-converted-to.pdf}}}  \nonumber \\
&+&  \vcenter{\hbox{\includegraphics[width=1.9in]{AFirstStepRR_4cVertex-eps-converted-to.pdf}}} \nonumber \\
&+& \left(  \vcenter{\hbox{\includegraphics[width=2.5in]{AFirstStepRR_IStart-eps-converted-to.pdf}}} + \vcenter{\hbox{\includegraphics[width=2.5in]{AFirstStepRR_AT_IStart-eps-converted-to.pdf}}} \right. \nonumber \\
&+& \left. \vcenter{\hbox{\includegraphics[width=2.5in]{AFirstStepRR_IIAPart-eps-converted-to.pdf}}} +  \vcenter{\hbox{\includegraphics[width=2.5in]{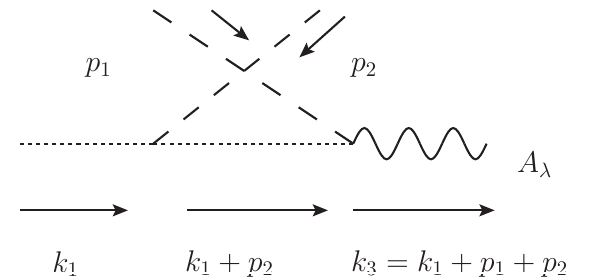}}}\right) \nonumber \\ 
&\cdot& (-i m_A), \label{RRReducedResult}
\end{eqnarray}
where we use $\doteq$ to replace the $=$, indicating that (\ref{RRReducedResult}) does not include the irrelevant terms in (\ref{FirstStepRR_ToBeCancelled}), which will finally be canceled out by other diagrams, or be attributed to some other terms in the Nielsen identity which have nothing to do with our current ghost chain extension processes.

If,  on the other hand, the (\ref{dAdXi}) encounters two $I$'s on one side,  we can sum over these following diagrams
\begin{eqnarray}
\vcenter{\hbox{\includegraphics[width=1.9in]{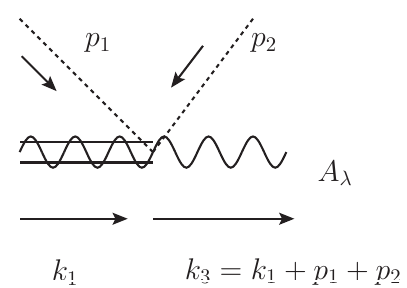}}} &=& 2 i g^2 [k_3^{\mu} - (p_1 + p_2)^{\mu}],  \label{AFirstStepII_4V}\\
\vcenter{\hbox{\includegraphics[width=2.5in]{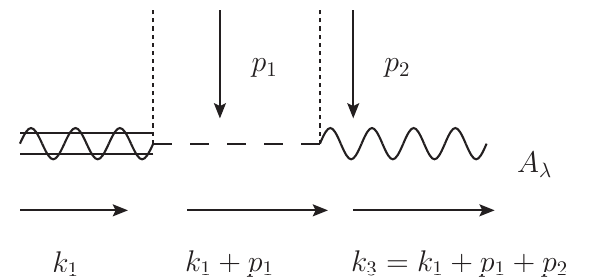}}} &=& 2 g [k_1 \cdot (-k_1-p_1)] \frac{i[2 g (k_1+p_1)^{\mu}]}{(k_1+p_1)^2 - m_R^2},  \label{AFirstStepII_s}\\
\vcenter{\hbox{\includegraphics[width=2.5in]{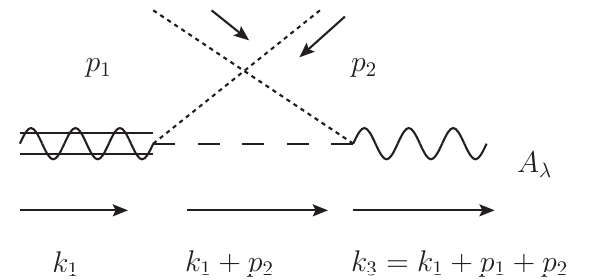}}} &=& 2 g [k_1 \cdot (-k_1-p_2)] \frac{i[2 g (k_1+p_2)^{\mu}]}{(k_1+p_2)^2 - m_R^2}. \label{AFirstStepII_t}
\end{eqnarray}
The final $\frac{-i}{k_3^2 - m_A^2} \left[g_{\mu \lambda} - \frac{k_{3 \mu} k_{3 \lambda}}{k_3^2-\xi m_A^2} (1-\xi) \right]$ propagator is again omitted for abbreviation. Notice, for example, in (\ref{AFirstStepII_s}), that $2 k_1 \cdot (-k_1-p_1) = p_1^2 - k_1^2 -(k_1+ p_1)^2$, so it is then reduced to
\begin{eqnarray}
& & 2 g [k_1 \cdot (-k_1-p_1)] \frac{i[2 g (k_1+p_1)^{\mu}]}{(k_1+p_1)^2 - m_R^2} \nonumber \\
&=& - i g (2 g) (k_1+p_1)^{\mu}  + g \frac{-i m_R^2 }{(k_1+p_1)^2 - m_R^2}  2 g (k_1+p_1)^{\mu} \nonumber \\
&-& g (k_1^2 - \xi m_A^2) \frac{i}{(k_1+p_1)^2 - m_R^2} 2 g (k_1+p_1)^{\mu} \nonumber \\
&+&  g (p_1^2 - \xi m_A^2) \frac{i}{(k_1+p_1)^2 - m_R^2} 2 g (k_1+p_1)^{\mu}. \label{FirstStepII_ToBeCancelled}
\end{eqnarray}
Again, the last term is destined to be canceled by some other diagrams, and the third term will contribute to  $\langle g \xi RI \rangle_{\text{1PI}}$ so we leave this for later discussion. This is similar to the case in (\ref{FirstStepRR_ToBeCancelled}).  The first term cancels the $p_1^{\mu}$ term in (\ref{AFirstStepII_4V}), and further cancels out the $k_3^{\mu}$ term. If we take (\ref{AFirstStepII_t}) into account, we will also kill the $p_2^{\mu}$ term in (\ref{AFirstStepII_4V}) and acquire an opposite $-2 i g^2 (k_3^{\mu})$ term. This will induce a ghost-ghost-$I$-$I$ vertex as before. The second term of (\ref{FirstStepII_ToBeCancelled}) is subtle. Notice the structure of the diagram. Tracing from $k_1$ to $p_1$ gives an alternate possibility to generate a new ghost chain, and this term actually corresponds to it, so here we do not have to reproduce the details. In fact, One can figure out that if $p_1$ further connects to a vector and a $R$ propagator, the first term in (\ref{FirstStepII_ToBeCancelled}) exactly corresponds to the second term in (\ref{FirstStepRR_ToBeCancelled}), giving a practical example of how to cancel this.

Finally, we have
\begin{eqnarray}
& & \vcenter{\hbox{\includegraphics[width=2.5in]{AFirstStepII-eps-converted-to.pdf}}} +   \vcenter{\hbox{\includegraphics[width=2.5in]{AFirstStepII_AT-eps-converted-to.pdf}}} \nonumber \\
&+&   \vcenter{\hbox{\includegraphics[width=1.9in]{AFirstStepII_4Vertex-eps-converted-to.pdf}}} \doteq \vcenter{\hbox{\includegraphics[width=1.9in]{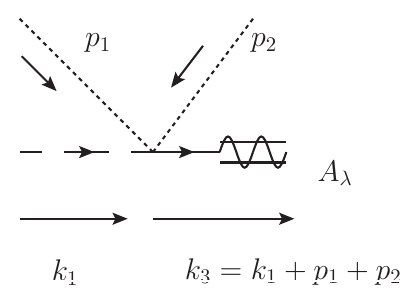}}}. \label{ccII}
\end{eqnarray}
Again ``$\doteq$'' implies that we tentatively neglect the terms which are irrelevant to our current ghost chain. These terms might be canceled by other diagrams, or contribute to other ghost chain generations. 

However, the above discussions depend on the assumption that both the (\ref{AFirstStepII_s}) and (\ref{AFirstStepII_t}) diagrams exist as parts of the 1-particle-irreducible (1PI) diagrams. Sometimes, one of them does not exist. For example, without loss of generality, if the $R$-propagator in (\ref{AFirstStepII_s}) separates the two parts of the diagram without other connections, then (\ref{AFirstStepII_s}) is not a 1PI and the $k_1+p_2$ propagator in (\ref{AFirstStepII_t}) becomes the common line as the waist of a gourd. Therefore we have to introduce the following diagram
\begin{eqnarray}
& & \vcenter{\hbox{\includegraphics[width=2.5in]{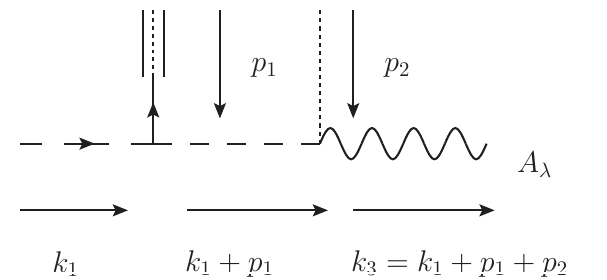}}} = (-2 i g^2 \xi v) (-i m_A) \frac{i[2 g (k_1+p_1)^{\mu}]}{(k_1+p_1)^2 - m_R^2} \nonumber \\
&=& -i g (2 g)(k_1+p_1)^{\mu}  \times \frac{i 2 g \xi v}{(k_1+p_1)^2 - m_R^2} (-i m_A)\nonumber \\
&\doteq&\frac{i 2 g \xi v}{(k_1+p_1)^2 - m_R^2}(-i m_A)  \vcenter{\hbox{\includegraphics[width=2.5in]{AFirstStepII-eps-converted-to.pdf}}}, \label{AFirstStepII}
\end{eqnarray}
This $\doteq$ is a little bit unintelligible, since we substitute the $-i g (2 g)(k_1 + p_1)^{\mu}$ with a much more complicated diagram. In fact, notice that it mimics with the first term of (\ref{FirstStepII_ToBeCancelled}) corresponding to (\ref{AFirstStepII_s}). The rest term of (\ref{FirstStepII_ToBeCancelled}) are again finally canceled out by other diagrams through the same tricks as the processes to prove the Ward identity diagrammatically. Therefore, in this case, (\ref{ccII}) should be modified into
\begin{eqnarray}
& & \vcenter{\hbox{\includegraphics[width=2.5in]{AFirstStepII_AT-eps-converted-to.pdf}}} +   \vcenter{\hbox{\includegraphics[width=1.9in]{AFirstStepII_4Vertex-eps-converted-to.pdf}}} \nonumber \\
&\doteq& \vcenter{\hbox{\includegraphics[width=1.9in]{AFirstStepII_4cVertex-eps-converted-to.pdf}}} - \frac{(k_1+p_1)^2 - m_R^2}{(-i m_A) (2 \xi g v i)}  \vcenter{\hbox{\includegraphics[width=2.5in]{AFirstStepII_Separation-eps-converted-to.pdf}}}. \label{ccII_Fixed}
\end{eqnarray}
Here we applied (\ref{AFirstStepII}) by multiplying it with the reciprocal of the factor $\frac{2 i g \xi v (-i m_A)}{(k_1+p_1)^2-m_R^2}$. The last term of (\ref{ccII_Fixed}) can be understood as the direct multiplication of two diagrams because the $(k_1+p_1)^2-m_R^2$ kills the propagator. This is exactly one example of the righted-hand side of the (\ref{NielsenRelying}), in which the propagator $k_1+p_1$ separated the bulk $\frac{\delta \Gamma}{\delta R}$ and the $C$-part.

Now we are going to prolong the ghost chain. The basic idea is to repeat the previous processes successively. We firstly assert that for every collection of the propagator chains started with (\ref{HalfA}) and ended with a vector boson, with a specific number of $R$ propagators accessing into it, it will finally be reduced into a collection of ghost chains ended with (\ref{HalfCHalfA}), and another collection of vector-Goldstone chains started with the $I$ and ended with a vector boson. That is to say,
\begin{eqnarray}
& & \sum_{\text{All possible connections}} \vcenter{\hbox{\includegraphics[width=3.2in]{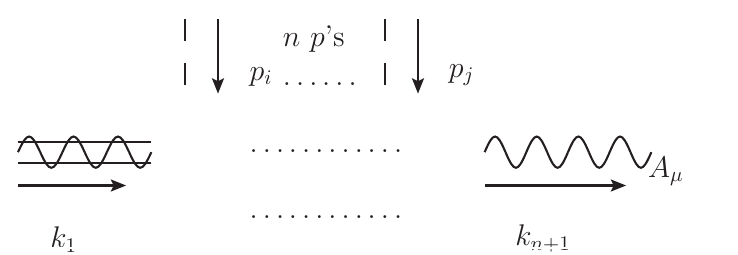}}} \nonumber \\
&\doteq& \sum_{\text{All possible connections}} \vcenter{\hbox{\includegraphics[width=3.2in]{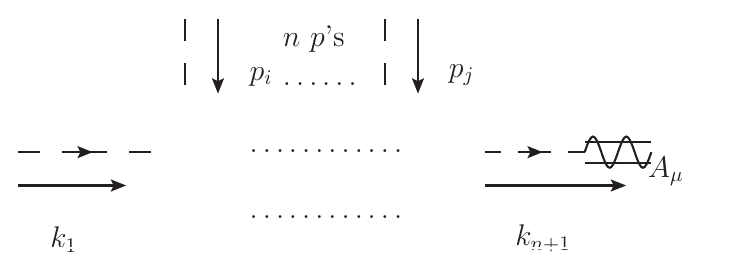}}} \nonumber \\
&+& (-i m_A)  \sum_{\text{All possible connections}} \ \vcenter{\hbox{\includegraphics[width=3.2in]{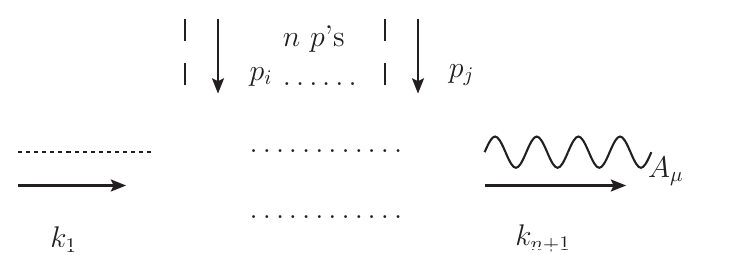}}}. \label{ABeginAFinalReduction}
\end{eqnarray}

We have already proved the first two steps, and we are going to extend this through the complete induction method. To achieve this, let us add an additional vector propagator at the end of (\ref{ABeginAFinalReduction}). If one additional $R$ is emitted,
\begin{eqnarray}
& & \sum_{\text{All possible connections}}  \vcenter{\hbox{\includegraphics[width=4.2in]{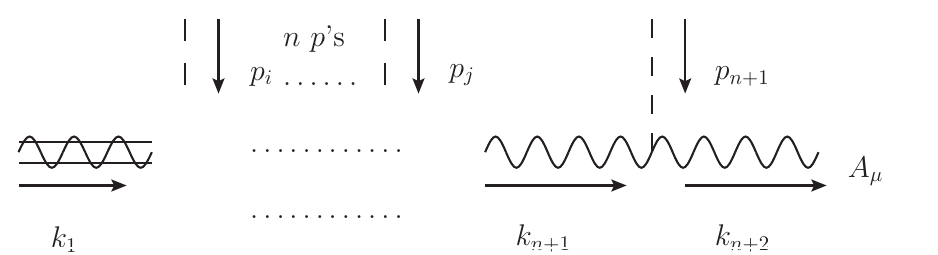}}} \nonumber \\
&\doteq& \sum_{\text{All possible connections}} \vcenter{\hbox{\includegraphics[width=4.2in]{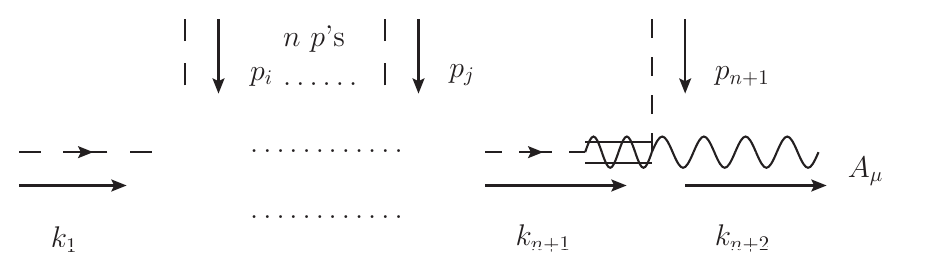}}} \nonumber \\
&+& (-i m_A)  \sum_{\text{All possible connections}} \ \vcenter{\hbox{\includegraphics[width=4.2in]{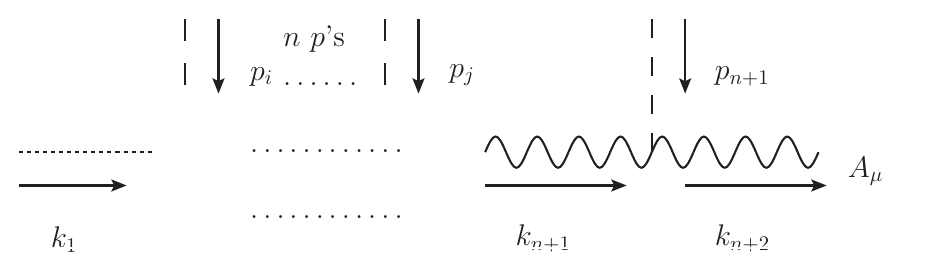}}}. \label{AddRAOneStep}
\end{eqnarray}
Following (\ref{FirstStepR_Decompose}), the first term of (\ref{AddRAOneStep}) can be reduced to
\begin{eqnarray}
& & \sum_{\text{All possible connections}} \vcenter{\hbox{\includegraphics[width=4.2in]{cBeginAFinal_AddRA-eps-converted-to.pdf}}} \nonumber \\
&\doteq& \sum_{\text{All possible connections}} \vcenter{\hbox{\includegraphics[width=4.2in]{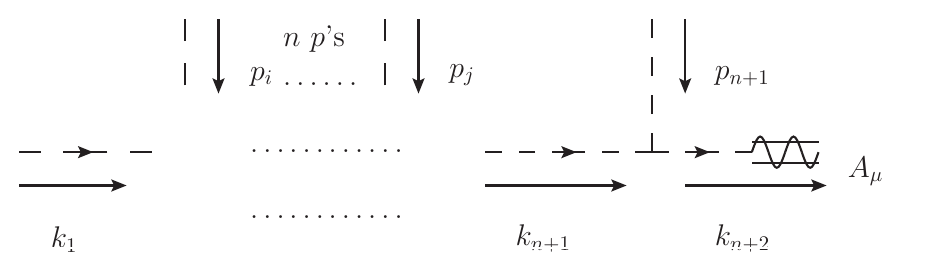}}} \nonumber \\
&+& \sum_{\text{All possible connections}} \vcenter{\hbox{\includegraphics[width=4.2in]{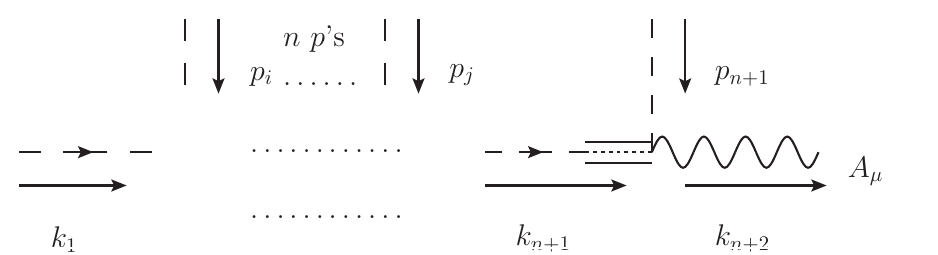}}}. \label{AddRAOneStep_HalfCHalfAReduced}
\end{eqnarray}
The $(-i m_A)$ factor have been absorbed into the half-ghost half-$I$ propagators defined in (\ref{HalfCHalfIPropagator}). The destination of the second term in (\ref{AddRAOneStep_HalfCHalfAReduced}) depends on the details before the $k_{n+1}$ propagators. For example,  if the second term of (\ref{AddRAOneStep_HalfCHalfAReduced}) ends up like (\ref{FirstStepRR_Cs})+(\ref{FirstStepRR_Ct}),  then there must exist diagrams ended up with patterns like (\ref{FirstStepRR_Is})+(\ref{FirstStepRR_It})+(\ref{FirstStepRR_4VertexResult}), which are not included in (\ref{AddRAOneStep}) because their $k_{n+1}$ propagator is not a vector boson. However, they start with a vector propagator and are with their $k_{n}$ propagator being a vector boson, so these segments have been manipulated similarly to (\ref{ABeginAFinalReduction}). By picking up such contributions, we arrive at diagrams ending up with patterns like (\ref{FirstStepRR_IIV}), with the half-ghost half-$I$ propagator moves backward by one propagator.  The sketched processes are depicted as
\begin{eqnarray}
& & \sum_{\text{All possible connections}} \vcenter{\hbox{\includegraphics[width=3.1in]{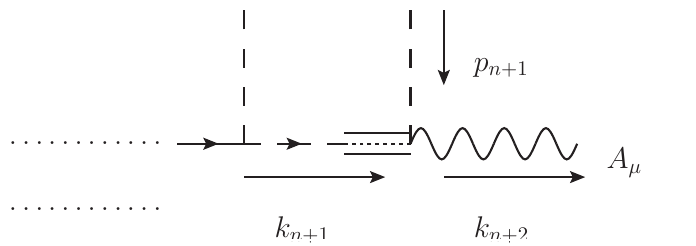}}} \nonumber \\
&+& \sum_{\text{All possible connections}} \vcenter{\hbox{\includegraphics[width=3.1in]{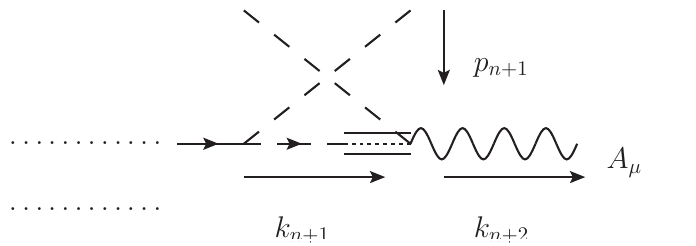}}} \nonumber \\
&+& \sum_{\text{All possible connections}} \vcenter{\hbox{\includegraphics[width=3.1in]{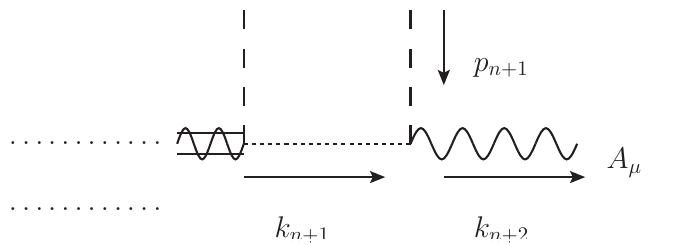}}} \nonumber \\
&+& \sum_{\text{All possible connections}} \vcenter{\hbox{\includegraphics[width=3.1in]{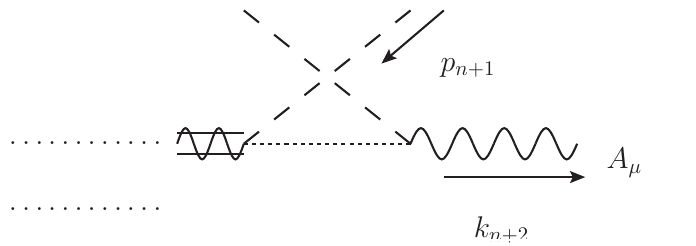}}}  \nonumber \\
&+& \sum_{\text{All possible connections}} \vcenter{\hbox{\includegraphics[width=3.1in]{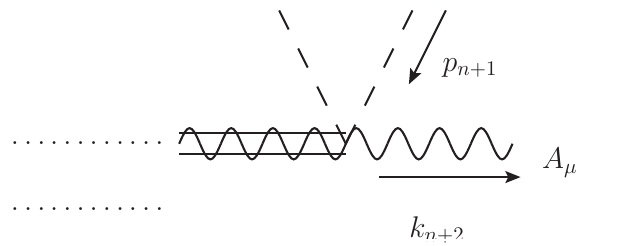}}} \nonumber \\
& \doteq& \sum_{\text{All possible connections}} \vcenter{\hbox{\includegraphics[width=3.1in]{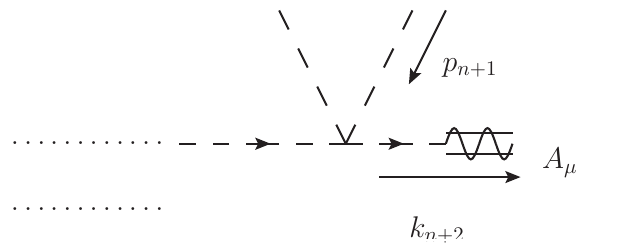}}} \nonumber \\
&+& \sum_{\text{All possible connections}} \vcenter{\hbox{\includegraphics[width=3.1in]{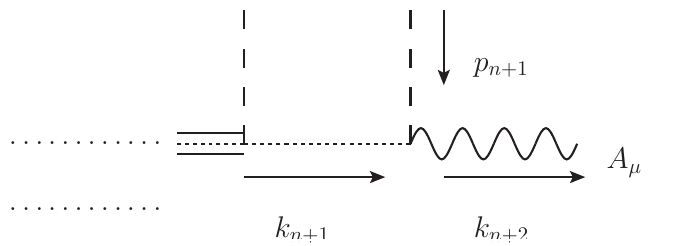}}} \nonumber \\
&+& \sum_{\text{All possible connections}} \vcenter{\hbox{\includegraphics[width=3.1in]{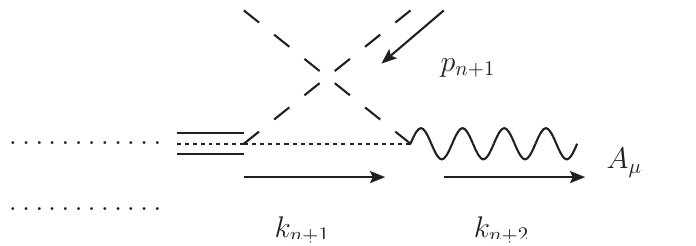}}}.  \label{ExampleOfProlonging}
\end{eqnarray}
Such processes can be repeated recursively, so that the half-ghost half-$I$ propagator moves backward consecutively and finally transmute the first vector boson to an $I$ for us to acquire a series of diagrams with a line of $I$ propagators acting as the spine. This will be attributed to the second term of (\ref{ABeginAFinalReduction}). To see this, let us enumerate all the possibilities to carry the half-ghost half-$I$ propagators backwards.

The first example is when the half-ghost half-$I$ propagator only emits one $R$ before it,
\begin{eqnarray}
\vcenter{\hbox{\includegraphics[width=1.3in]{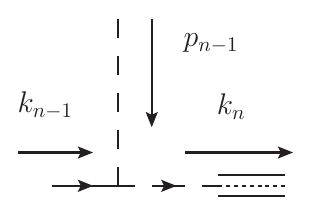}}} &=& (-2 i \xi g^2 v) \frac{i}{k_n^2 - \xi m_A^2}(-i m_A). \label{cIRI}
\end{eqnarray}
Notice that (\ref{ABeginAFinalReduction}) is only available for the diagrams ended with a vector boson. There must exist diagrams emitting exactly the same $R$ propagators, however ended with a series of $I$'s, ensuing shorter ghost chains while clinging to a series of $I$'s. Therefore, within these diagrams there must exist one with the following spare part,
\begin{eqnarray}
\vcenter{\hbox{\includegraphics[width=1.3in]{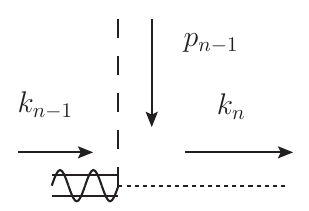}}} &=& 2 g(k_{n-1} \cdot p_{n-1})  \frac{i}{k_n^2 - \xi m_A^2}.  \label{AIRI}
\end{eqnarray}
$2 k_{n-1} \cdot p_{n-1} = k_n^2 - p_{n-1}^2 - k_{n-1}^2$, and concentrate on the terms that no propagator pole is canceled, so we replace $2 k_{n-1} \cdot p_{n-1}$ with $\xi m_A^2 - m_R^2 - \xi m_A^2$,. Then we have
\begin{eqnarray}
& & (\ref{cIRI})+(\ref{AIRI}) \doteq (-2 i \xi g^2 v - 2 \lambda v) \frac{i}{k_n^2 - \xi m_A^2}(-i m_A) \nonumber \\
&=& \vcenter{\hbox{\includegraphics[width=1.3in]{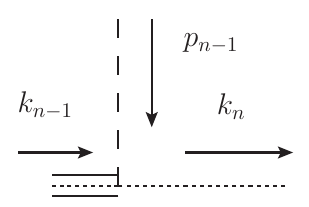}}}, \label{OneUnitBackward}
\end{eqnarray}
thus moves the half-gost half-$I$ propagator backward by one propagator.

If we pick up the terms in (\ref{AIRI}) that kills some of the poles of the propagators,  we can encounter the $I^2 R^2$ vertex
\begin{eqnarray}
& & \vcenter{\hbox{\includegraphics[width=2.5in]{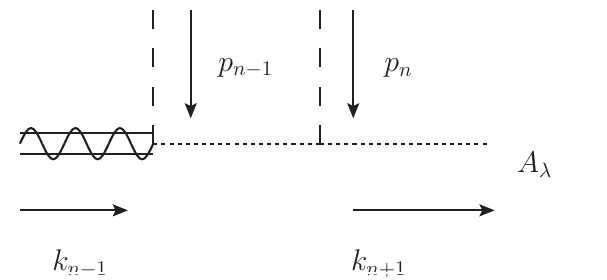}}} + \vcenter{\hbox{\includegraphics[width=2.5in]{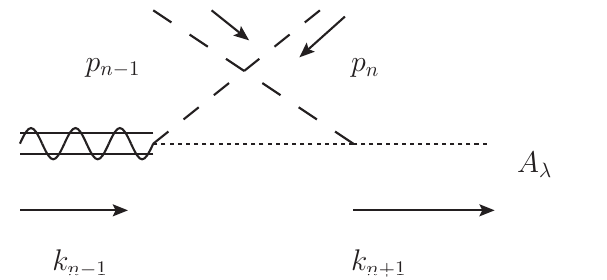}}} \nonumber  \\
&\rightarrow& 2 i g [-2 i v ( \lambda + g^2 \xi)], \label{ARIR} \\
& & \vcenter{\hbox{\includegraphics[width=2.5in]{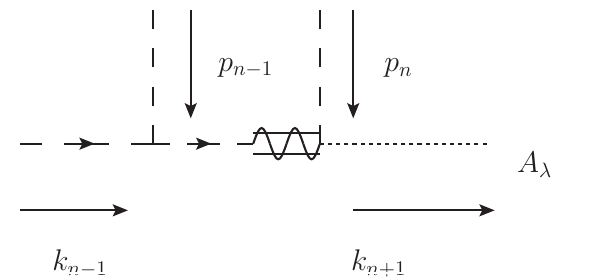}}} + \vcenter{\hbox{\includegraphics[width=2.5in]{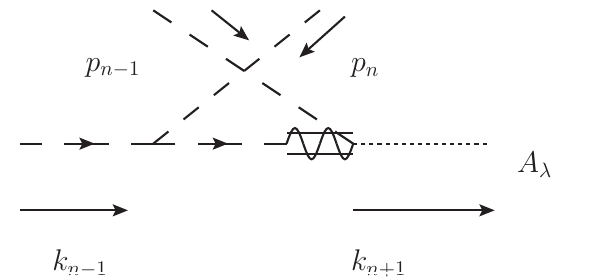}}} ] \nonumber \\
&\rightarrow& (-2 i g^2 \xi v)  [-2 i g],  \label{cRIR}  \\
& & \vcenter{\hbox{\includegraphics[width=1.5in]{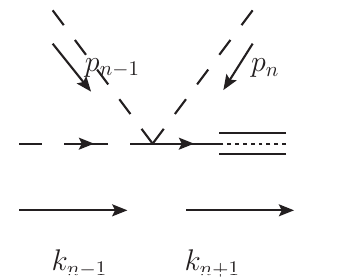}}} =  (-2 i g^2 \xi )  (-i m_A) \label{ccRR}
\end{eqnarray}
Here we only preserve the factors to kill the propagators in the middle while neglect other terms. Sum over (\ref{ARIR})+(\ref{cRIR})+(\ref{ccRR}) we acquire
\begin{eqnarray}
& & (-i m_A) (-2 i)(\lambda + g^2 \xi) +(-i g) (6 \lambda v i) \\
&=&\vcenter{\hbox{\includegraphics[width=1.5in]{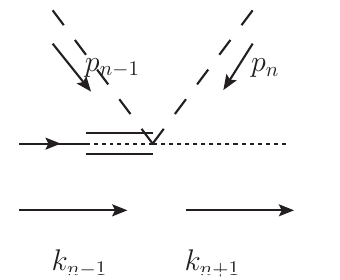}}} -  \frac{[(p_{n-1}+p_n)^2 - m_R^2]}{(-i m_A) (2 \xi g v i)} \cdot \vcenter{\hbox{\includegraphics[width=1.5in]{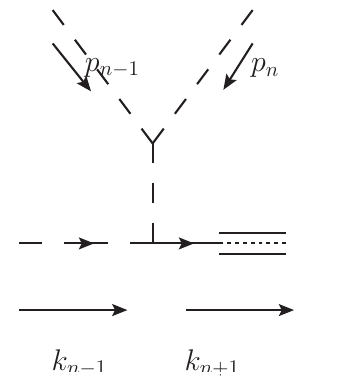}}} \label{cIIRR_ccI_RRR}
\end{eqnarray}
The second term of (\ref{cIIRR_ccI_RRR}) might be canceled by other terms as the situation of the diagrammatic proof of the Ward-Takahashi identities. However, if $k_{n-1}$ and $k_{n+1}$ propagators appear to clamp the waist of the gourd, this diagram will no longer be a part  of the 1PI-diagram. Therefore such diagrams will drop down and remain during the prolonging processes. We will see that this is exactly the separation process between the bulk $\frac{\delta \Gamma}{\delta R}$ and the $C$-part of the (\ref{NielsenRelying}).

Another example is when an $R^2 I^2$ vertex is prefixed by emitting another $R$ propagator,
\begin{eqnarray}
\vcenter{\hbox{\includegraphics[width=2.3in]{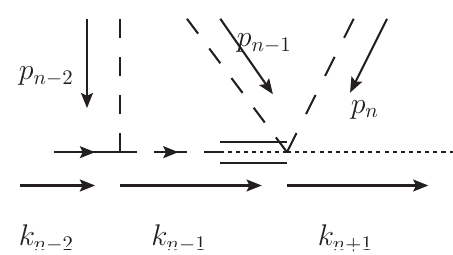}}} &=& (-2 i g^2 v \xi) \frac{i (-i m_A)}{k_{n-1}^2 - \xi m_A^2} [(-2 i )(\lambda + \xi g^2)], \\
\vcenter{\hbox{\includegraphics[width=2.3in]{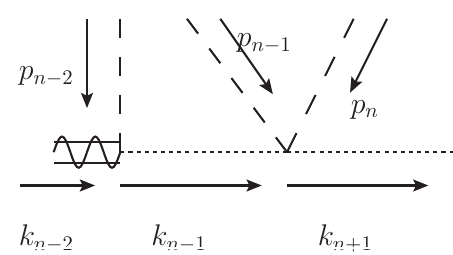}}} &=& 2 g (k_{n-2} \cdot p_{n-2})  \frac{i}{k_{n-1}^2 - \xi m_A^2} [(-2 i )(\lambda + \xi g^2)]. \label{A_AIIRR}
\end{eqnarray}
These two diagrams induce a 
\begin{eqnarray}
\vcenter{\hbox{\includegraphics[width=2.3in]{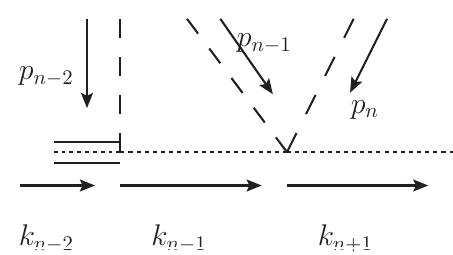}}}
\end{eqnarray}
according to (\ref{OneUnitBackward}). However, the pole of the $k_{n-1}$ propagator can be killed since $2 p_{n-2} \cdot k_{n-2} = k_{n-1}^2 - p_{n-2}^2 - k_{n-2}^2$ and $k_{n-1}^2$ can induce a term $k_{n-1}^2 - \xi m_A^2$ to cancel the pole.  Coordinated with the following diagram,
\begin{eqnarray}
\vcenter{\hbox{\includegraphics[width=2.3in]{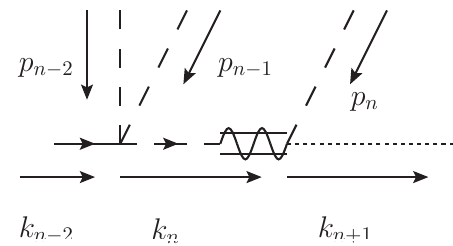}}} &=& (- 2 i g^2 \xi) \frac{i}{k_{n}^2 - \xi m_A^2} (2 g) p_{n} \cdot k_{n-1}, \label{A_AIIRR_t}
\end{eqnarray}
where $2 p_n \cdot k_{n} = k_{n+1}^2 - p_n^2 - k_{n-1}^2$, generating a term to kill the $k_{n-1}^2 = \xi m_A^2$ pole.  Finally,
\begin{eqnarray}
& & \vcenter{\hbox{\includegraphics[width=2.3in]{A_AIIRR-eps-converted-to.pdf}}}  + \vcenter{\hbox{\includegraphics[width=2.3in]{A_AIIRR_t-eps-converted-to.pdf}}} \nonumber \\
&\rightarrow& i g (-2 i \lambda) = -\frac{1}{3} \frac{(p_{n-2}+p_{n-1}+p_{n})^2 - m_R^2}{(2 g \xi v i) (-i m_A)} \cdot \vcenter{\hbox{\includegraphics[width=1.5in]{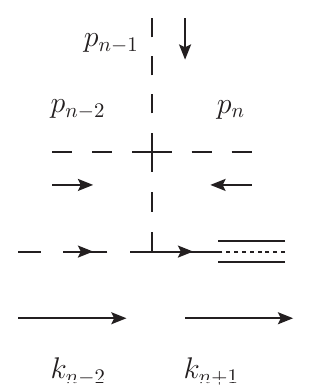}}}. \label{ccI_RRRR}
\end{eqnarray}
Again, this term will be canceled if $k_{n-2}$ and $k_{n+1}$ do not clamp the waist, and will be dropped down to form the righted-hand side of the (\ref{NielsenRelying}) in the counter case.  Notice that exchanging $p_{n-2}$, $p_{n-1}$ and $p_{n}$ in (\ref{A_AIIRR}) and (\ref{A_AIIRR_t}) gives three different patterns of momentum flows, thus form a factor of $3$ to supplement the $\frac{1}{3}$ in (\ref{ccI_RRRR}).


The prolonging progresses might also encounter an $I^4$ vertex defined in (\ref{RRRR_IIII_Vertex}). An example can be
\begin{eqnarray}
\vcenter{\hbox{\includegraphics[width=2.3in]{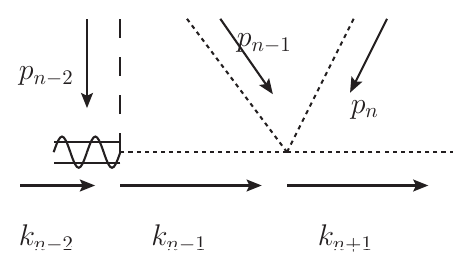}}}  &=& 2 g (p_{n-2} \cdot k_{n-2}) \frac{i}{k_{n-1}^2 - \xi m_A^2} (-6 i \lambda). \label{A_IIII}
\end{eqnarray}
Because $2 p_{n-2} \cdot k_{n-2} = k_{n-1}^2 - p_{n-2}^2 - k_{n-2}^2$,
\begin{eqnarray}
 & & g (p_{n-2} \cdot k_{n-2}) \frac{i}{k_{n-1}^2 - \xi m_A^2} (-6 i \lambda) \nonumber \\
 &=& g (-6 i \lambda) - g (-6 i \lambda) m_R^2 \frac{i}{k_{n-1}^2 - \xi m_A^2}   \nonumber \\
 &  & - (p_{n-2}^2 - m_R^2) g \frac{i}{k_{n-1}^2 - \xi m_A^2} (-6 i \lambda) - (k_{n-2}^2 - \xi m_A^2) g \frac{i}{k_{n-1}^2 - \xi m_A^2} (-6 i \lambda). \label{IIII_Reduced}
\end{eqnarray}
Again, the last two terms cancels other propagators, which will be canceled by other diagrams, or can be ascribed to the results of the  tree-level $\frac{\delta \Gamma}{\delta R}$ times the loop level $C$-part. The second term can be depicted as
\begin{eqnarray}
\vcenter{\hbox{\includegraphics[width=2.3in]{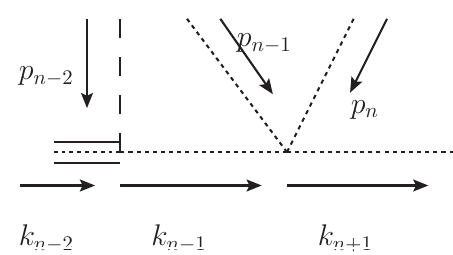}}},
\end{eqnarray}
and will participate the processes to move the half-ghost propagators backwards and backwards as we have described. It is the first term of (\ref{IIII_Reduced}) that we have to concern. To cancel this, we have to calculate the following two diagrams,
\begin{eqnarray}
\vcenter{\hbox{\includegraphics[width=2.0in]{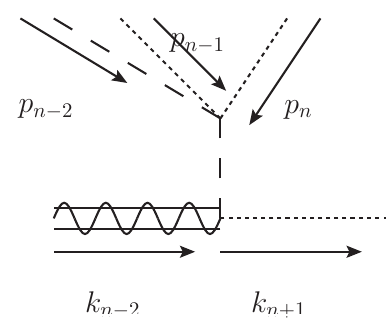}}} &=& 2 g   \frac{i[(p_{n-2} + p_{n-1}+p_n) \cdot k_{n-2}]}{(p_{n-2}+p_{n-1}+p_n)^2 - m_R^2} (-2 i) (\lambda + \xi g^2),  \label{IIIICancallation1}\\\vcenter{\hbox{\includegraphics[width=2.3in]{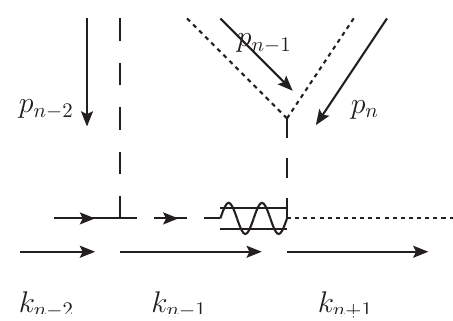}}} &=& 2 \xi g^2 i   \frac{i}{(p_{n-1}+p_n)^2 - m_R^2} (2 g ) [k_{n-1} \cdot (p_n+p_{n-1})].  \label{IIIICancellation2}
\end{eqnarray}
For (\ref{IIIICancallation1}), $2(p_{n-2} + p_{n-1}+p_n) \cdot k_{n-2} = k_{n+1}^2 - k_{n-2}^2 - (p_{n-2} + p_{n-1}+p_n)^2$, and the $- (p_{n-2} + p_{n-1}+p_n)^2$ term will lead to cancel the $( p_{n-2} + p_{n-1}+p_n)^2= m_R^2$ pole, thus contribute to a $ g (2 i )(\lambda + \xi g^2)$ term. For the (\ref{IIIICancellation2}), $2 k_{n-1} \cdot (p_n+p_{n-1}) = k_{n+1}^2 - (p_{n-1}+p_n)^2 - k_{n-1}^2$, and the $- (p_{n-1}+p_n)^2$ will lead to cancel the $( p_{n-1}+p_n)^2= m_R^2$ pole, leaving us a $g (-2 i) \xi g^2$. Summing over these two terms gives the result of $-2 i \lambda g$. If $p_{n-2}$, $p_{n-1}$, $p_{n}$ all belong to the $C$-part of the (\ref{NielsenRelying}), (\ref{IIIICancallation1}) and (\ref{IIIICancellation2}) will cancel $\frac{1}{3}$ of the first term in (\ref{IIII_Reduced}). Swapping the $k_{n+1}$ with $p_n$ or $p_{n-1}$ induces the remained $\frac{2}{3}$. This also means that whenever we encounter the $I^4$ vertex, there are three possibilities for us to select a route to extend the ghost chain through each $I$. We have to consider all of their contributions.

If, on the other hand, the $k_{n-1}$ propagator in (\ref{A_IIII}) is the common internal line (or the ``waist'') shared by both the bulk part and $C-$part of the (\ref{NielsenRelying}), either or both (\ref{IIIICancallation1}) and (\ref{IIIICancellation2}) might not exist due to the property of 1PIs.  Again, diagrams like
\begin{eqnarray}
-\frac{(p_{n-2}+p_{n-1}+p_{n})^2 - m_R^2}{(2 g \xi v i) (-i m_A)} \cdot \vcenter{\hbox{\includegraphics[width=2.0in]{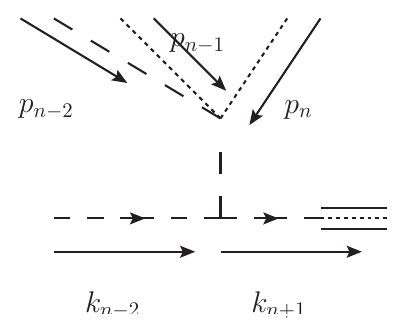}}} &=& i g (-2 i)(\lambda + \xi g^2),  \label{c_IIII1}\\
-\frac{(p_{n-1}+p_{n})^2 - m_R^2}{(2 g \xi v i) (-i m_A)} \cdot \vcenter{\hbox{\includegraphics[width=2.3in]{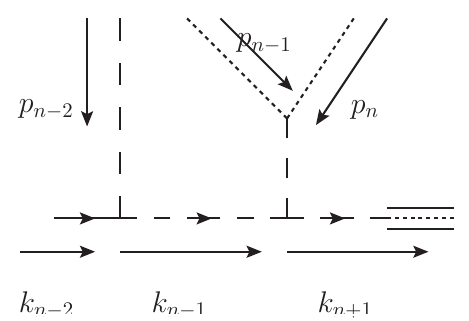}}} &=& -2 \xi g^2  (- g i ).  \label{c_IIII2}
\end{eqnarray}
can arise to form the separated bulk-part and $C$-part on the righted side of (\ref{NielsenRelying}).

Similarly, vector-vector-$I^2$ vertex in (\ref{AASS}) has to be concerned. Although this vertex get involved in (\ref{AFirstStepII_4V}), we have to point out that the ghost-chain might access from another direction,
\begin{eqnarray}
\vcenter{\hbox{\includegraphics[width=2.3in]{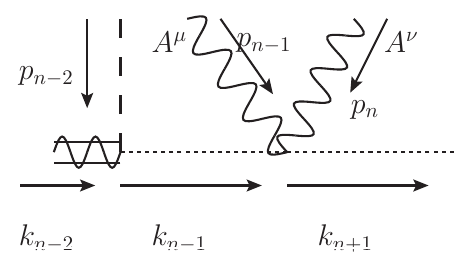}}} &=& 2 g (k_{n-2} \cdot p_{n-2}) \frac{i}{k_{n-1}^2 - \xi m_A^2} (2 i g^2 g^{\mu \nu}). \label{A_IIAA}
\end{eqnarray}
This diagram should accompany with at least the following diagrams
\begin{eqnarray}
& & \vcenter{\hbox{\includegraphics[width=2.0in]{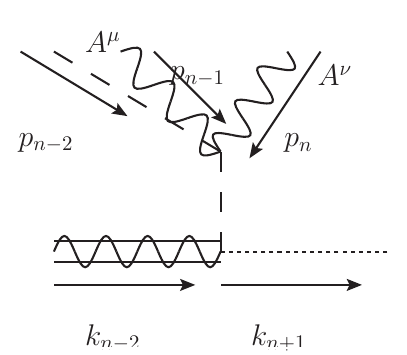}}} + 
\vcenter{\hbox{\includegraphics[width=2.3in]{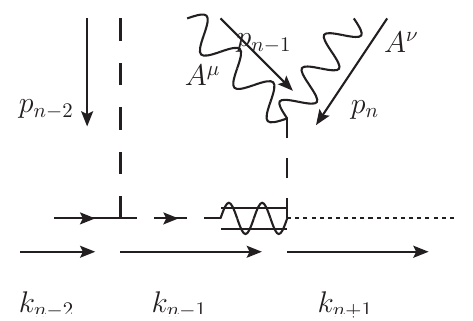}}}  \nonumber \\
&+& \vcenter{\hbox{\includegraphics[width=2.3in]{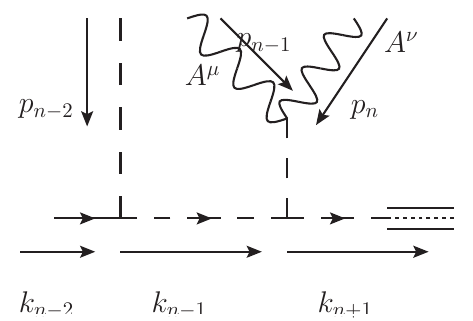}}} + \vcenter{\hbox{\includegraphics[width=2.3in]{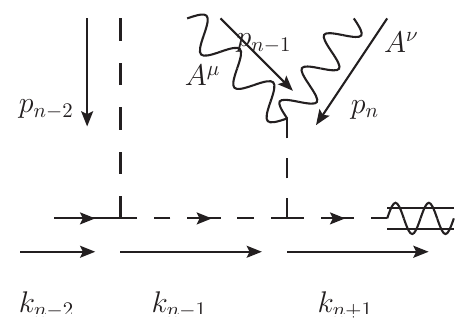}}} + \dots \label{A_IIAA_Various}
\end{eqnarray}
to prolong the ghost chain.  Within them the first and the second diagrams are crucial,
\begin{eqnarray}
\vcenter{\hbox{\includegraphics[width=2.0in]{A_IIAA_RAA-eps-converted-to.pdf}}}  &=& \frac{2 i g [k_{n-2} \cdot (p_{n-2}+p_{n-1}+p_{n})] }{(p_{n-2}+p_{n-1}+p_{n})^2 - m_R^2} 2 g^2 i g^{\mu \nu},  \label{A_IIAA_RAA}\\
\vcenter{\hbox{\includegraphics[width=2.3in]{A_IIAA_AA_ccI-eps-converted-to.pdf}}}  &=& \frac{i  (-2 i g^2 \xi) }{k_{n-1}^2 - \xi m_A^2}\frac{i  [2 g k_{n-1} \cdot (p_{n-1}+p_n)]  }{(p_{n-1}+p_n)^2 - m_R^2} (2 g^2 i v). \label{A_IIAA_AA_ccI}
\end{eqnarray}
For the (\ref{A_IIAA_RAA}), we have $k_{n-2} \cdot (p_{n-2}+p_{n-1}+p_{n}) = (k_{n-2} + p_{n-2}+p_{n-1}+p_{n})^2 - k_{n-2}^2 -  (p_{n-2}+p_{n-1}+p_{n})^2$, generating a term to kill the $(p_{n-2}+p_{n-1}+p_{n})^2 = m_R^2$ pole, and this will cancel with the corresponding term in (\ref{A_IIAA}) that $2 k_{n-2} \cdot p_{n-2} = k_{n-1}^2 - k_{n-2}^2 - p_{n-2}^2$ generates a $k_{n-1}^2 - \xi m_A^2$ term to kill the $k_{n-1}$ propagator. If, however, $k_{n-1}$ acts as the waist of the gourd, so all $p_{n-2}$, $p_{n-1}$, and $p_{n}$ are a part of the bulk $\frac{\delta \Gamma}{\delta R}$, (\ref{A_IIAA_RAA}) disappears for it is not an 1PI diagram and we have a supplement term
\begin{eqnarray}
& & \vcenter{\hbox{\includegraphics[width=2.3in]{A_IIAA-eps-converted-to.pdf}}} \rightarrow -g i (2 i g^2 g^{\mu \nu}) \nonumber \\ 
&=&- \frac{(p_{n-2}+p_{n-1}+p_{n})^2 - m_R^2}{(2 g \xi v i) (-i m_A)} \cdot \vcenter{\hbox{\includegraphics[width=2.0in]{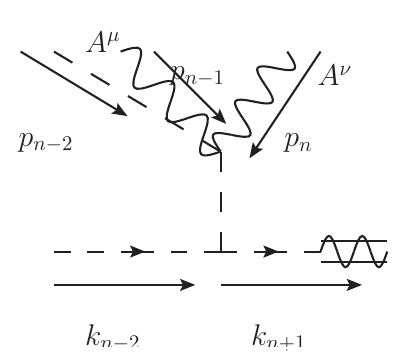}}}. \label{VVWaist1}
\end{eqnarray}
This term can be analyzed to be the multiplication of the bulk- and $I$- part of the (\ref{NielsenRelying}).

For the (\ref{A_IIAA_AA_ccI}), again, $2 k_{n-1} \cdot (p_{n-1}+p_n) = k_{n+1}^2 - k_{n-1}^2 - (p_{n-1}+p_n)^2$, generating the term to kill the $(p_{n-1}+p_n)^2 = m_R^2$ pole and to give rise to the term
\begin{eqnarray}
\vcenter{\hbox{\includegraphics[width=2.3in]{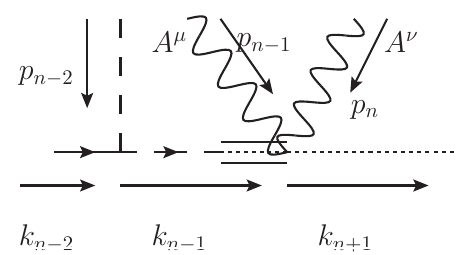}}}   &=& \frac{i  (-2 i g^2 \xi) }{k_{n-1}^2 - \xi m_A^2} (-i m_A) (2 g^2 i g^{\mu \mu}). \label{A_IIAA_AA_ccII}
\end{eqnarray}
Combined with the (\ref{A_IIAA}), one can further follow (\ref{OneUnitBackward}) to shift the half-ghost half-$I$ propagator backwards,
\begin{eqnarray}
 & &\vcenter{\hbox{\includegraphics[width=2.3in]{A_IIAA-eps-converted-to.pdf}}}   + \vcenter{\hbox{\includegraphics[width=2.3in]{A_IIAA_AA_ccII-eps-converted-to.pdf}}}  \nonumber \\
&\doteq& \vcenter{\hbox{\includegraphics[width=2.3in]{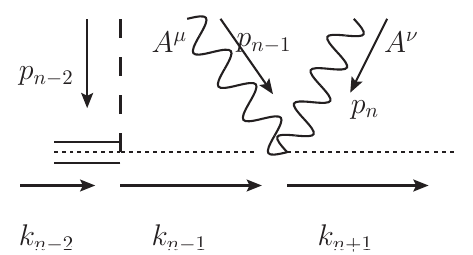}}}. \label{IIAA_Backward}
\end{eqnarray}
Thus the prolonging processes continue.

However, if the vector-vector-$I^2$ vertex inside (\ref{A_IIAA}) acts as the waist of the gourd, and the two vectors belong to the bulk $\frac{\delta \Gamma}{\delta R}$ part, diagrams in (\ref{A_IIAA_Various}) become absent because they are no longer 1PI's.  To let the processes in (\ref{IIAA_Backward}) continue, we can rewrite this into
\begin{eqnarray}
 & &\vcenter{\hbox{\includegraphics[width=2.3in]{A_IIAA-eps-converted-to.pdf}}}   \nonumber \\
&\doteq& \vcenter{\hbox{\includegraphics[width=2.3in]{A_IIAA_AA_cIII-eps-converted-to.pdf}}} - \vcenter{\hbox{\includegraphics[width=2.3in]{A_IIAA_AA_ccII-eps-converted-to.pdf}}} . \label{IIAA_C_Part}
\end{eqnarray}
Notice that
\begin{eqnarray}
\vcenter{\hbox{\includegraphics[width=2.3in]{A_IIAA_AA_ccc-eps-converted-to.pdf}}} =  \frac{i (-2 i g^2 \xi)}{k_{n-1}^2 - \xi m_A^2} \frac{i (-2 i g^2 \xi)  }{(p_{n-1}+p_n)^2 - m_R^2}  (2 i g^2 v g^{\mu \nu}) (-i m_A).
\end{eqnarray}
Compared with (\ref{A_IIAA_AA_ccII}), we have
\begin{eqnarray}
& & \vcenter{\hbox{\includegraphics[width=2.3in]{A_IIAA-eps-converted-to.pdf}}}   \doteq  \vcenter{\hbox{\includegraphics[width=2.3in]{A_IIAA_AA_cIII-eps-converted-to.pdf}}} \nonumber \\
&-& \frac{(p_{n-1}+p_{n})^2 - m_R^2}{(2 g \xi v i) (-i m_A)} \cdot \vcenter{\hbox{\includegraphics[width=2.3in]{A_IIAA_AA_ccc-eps-converted-to.pdf}}}  . \label{IIAA_C_Part}
\end{eqnarray}
The last term can also be explained as the multiplication of the two parts of the right-hand side of (\ref{NielsenRelying}).

\subsection{Separation of the $C$-part with the bulk $\frac{\delta \Gamma}{\delta \phi}$ part and the cancellation of the remained terms}

After we have anatomized out the spine of the ghost chain, we can now follow the direction that the chain elongate until it hits the end of the vector-$I$ chain that one select. During the prolonging processes, when the ghost chain hits the waist structure, the corresponding terms contributing to the right-hand side of (\ref{NielsenRelying}) arise. To enumerate the possibilities, we have to classify the waist types to calculate them respectively. There are six types of the waists, depending on the properties of the closest propagators on the bulk-part that touches the $C$-part, which are $R$-type, vector-vector type,  vector-$I$ type, $I$-$I$ type, ghost type, and the fermionic type.

The $R$-type waists had been discussed and enumerated in (\ref{cIIRR_ccI_RRR}) and (\ref{ccI_RRRR}). The vector-vector waists were discussed in (\ref{VVWaist1}) and (\ref{IIAA_C_Part}). The vector-$I$ type waists were displayed in (\ref{ccII_Fixed}),  and the $I$-$I$ type was illustrated in (\ref{c_IIII1}) and (\ref{c_IIII2}).  We also have to note that the bulk part might be a tree-level diagram, and can be isolated during the prolonging processes. As we have pointed out after (\ref{FirstStepRR_ToBeCancelled}),  its second term indicates the tree-level $(-\partial^2 - m_R^2) R \in \frac{\partial \Gamma}{\partial R}$ in the bulk if the $R$-propagator is external, and one can compare the coefficient with the one predicted by (\ref{NielsenRelying}).   (\ref{cIIRR_ccI_RRR}) and (\ref{ccI_RRRR}) can also be interpreted as the tree-level $\frac{\delta R^3}{\delta R}$ and $\frac{\delta R^4}{\delta R}$ terms adhering to the bulk-part if the corresponding $R$-propagators are external.

In this subsection, we concentrate on the ghost type and the fermionic type waists.

Incredibly the ghost type of waist originate from the vector-$I$ chains during the prolonging processes of the ghost chain. Following (\ref{ABeginAFinalReduction}),  the prolonging processes can extend from one point of this common line, and finalize to the other point,
\begin{eqnarray}
& & \vcenter{\hbox{\includegraphics[width=1.8in]{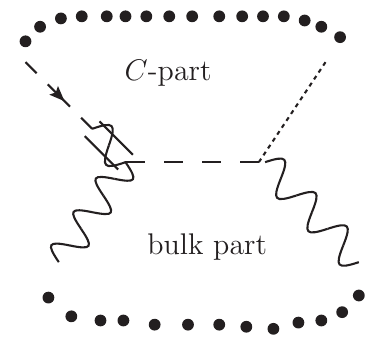}}} \doteq (-i m_A) \vcenter{\hbox{\includegraphics[width=1.8in]{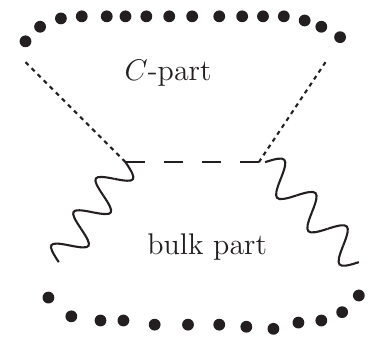}}} \nonumber \\
&+& \vcenter{\hbox{\includegraphics[width=1.8in]{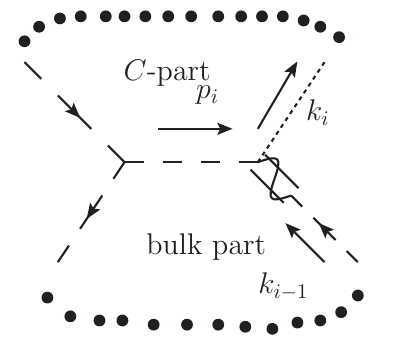}}} +\vcenter{\hbox{\includegraphics[width=1.8in]{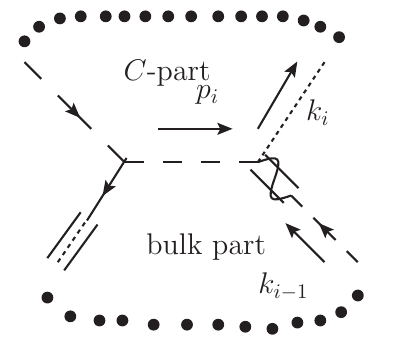}}} .
\end{eqnarray}
For the second  and the third term, around the $p_i$ propagator, the part of the diagram is expressed as,
\begin{eqnarray}
\vcenter{\hbox{\includegraphics[width=1.8in]{AAType_cPart-eps-converted-to.pdf}}} + \vcenter{\hbox{\includegraphics[width=1.8in]{AAType_cIPart-eps-converted-to.pdf}}} \supset 2 g (k_{i-1} \cdot p_i) \frac{i}{p_i^2 - m_R^2} ( 2 i g^2 \xi).
\end{eqnarray}
Again, $2 k_{i-1} \cdot p_i = k_{i}^2 - p_i^2 - k_{i-1}^2$, so
\begin{eqnarray}
& & 2 g (k_{i-1} \cdot p_i) \frac{i}{p_i^2 - m_R^2} ( 2 i g^2 \xi) \nonumber \\
& =  &- g i (2 i g^2 \xi) - g (k_{i-1}^2 - \xi m_A^2) \frac{i}{p_i^2 - m_R^2} ( 2 i g^2 \xi)+ g (k_i^2 - \xi m_A^2)\frac{i}{p_i^2 - m_R^2} ( 2 i g^2 \xi) \nonumber \\
& & + g (-m_R^2) g \frac{i}{p_i^2 - m_R^2} ( 2 i g^2 \xi). \label{AATypeGourd}
\end{eqnarray}
Terms other than the first term in (\ref{AATypeGourd}) will participate in the subsequent prolonging processes of the ghost-chain, so we do not care them. The first term can be re-expressed as
\begin{eqnarray}
- g i (2 i g^2 \xi) = - \frac{p_s^2 - m_R^2}{(2 g \xi v i) (-i m_A)} \cdot \vcenter{\hbox{\includegraphics[width=1.8in]{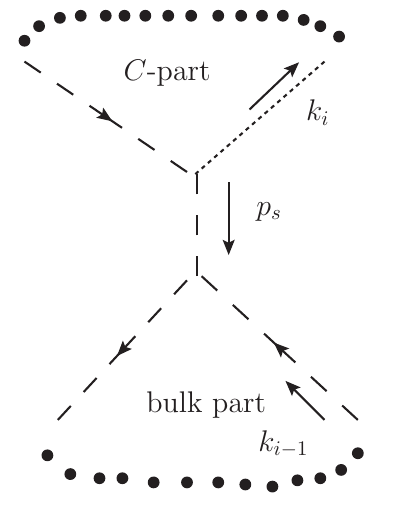}}}, \label{AAType_cIsolated}
\end{eqnarray}
which is exactly the ghost-type waist separating the bulk and the $C$-part of the diagrams described at the right-hand side of (\ref{NielsenRelying}).

For the fermionic type of waists, the $I$ or vector boson can insert into a fermionic loop to shift the $\psi_{1,2}$ to each other, so only even numbers of $I$/$A^{\mu}$ can connect with the closed fermionic loop. When a vector boson is connecting a pair of fermionic lines, we have,
\begin{eqnarray}
& & \vcenter{\hbox{\includegraphics[width=1.6in]{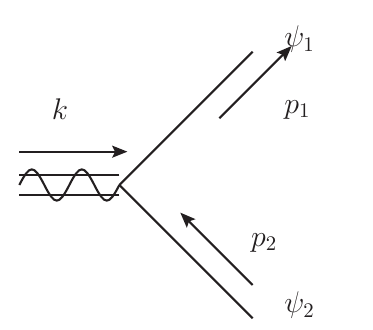}}} =  -Q_{\psi} g k_{\mu} \gamma^{\mu} \nonumber \\
&=& -Q_{\psi} g (p_{1 \mu} \gamma^{\mu}  - m_1) + Q_{\psi} g ( p_{2 \mu}\gamma^{\mu} - m_2) + Q_{\psi} g (m_1-m_2). \label{AToI}
\end{eqnarray}
The first two terms cancel the poles of the $\psi_1$ and $\psi_2$ propagators respectively, and will finally be canceled by other diagrams in which the $k$ propagator migrates along the fermionic loop\cite{Peskin:1995ev}. For the third term, recall from (\ref{m1m2}) that $m_2-m_1 = 2 \delta m = 2 y v$,  and $Q_{\psi}= -\frac{1}{2}$, so
\begin{eqnarray}
Q_{\psi} g (m_1-m_2) = y g v = -i m_A (i y) =  \vcenter{\hbox{\includegraphics[width=1.6in]{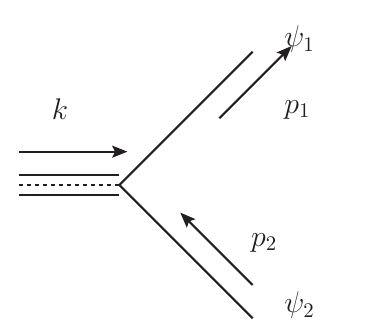}}}.
\end{eqnarray}
This will participate the processes like (\ref{cIRI}), (\ref{AIRI}) and (\ref{OneUnitBackward}) to move the half-ghost half-$I$ propagator backwards, leaving us an $I$-chain for further processes.

Finally, when the vector propagator migrates along the fermionic line, it will finally encounter an $I$ propagator as its neighbor. For example,
\begin{eqnarray}
& &  \vcenter{\hbox{\includegraphics[width=1.8in]{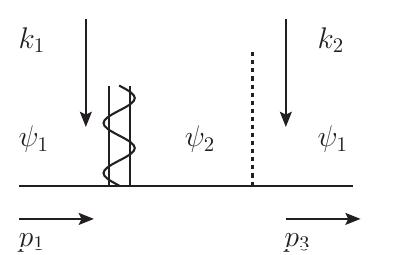}}} +  \vcenter{\hbox{\includegraphics[width=1.8in]{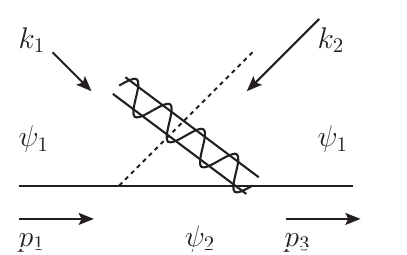}}} \nonumber \\
 &\doteq& \vcenter{\hbox{\includegraphics[width=1.8in]{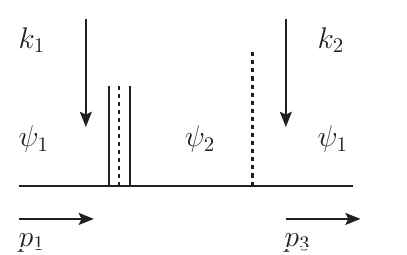}}} +  \vcenter{\hbox{\includegraphics[width=1.8in]{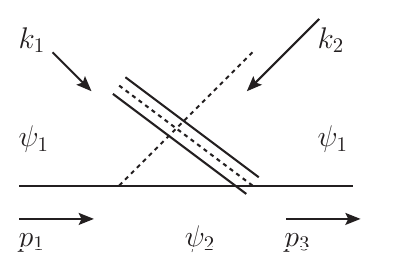}}} \nonumber \\
 &-&  \frac{(k_1+k_2)^2 - m_R^2}{(2 g \xi v i) (-i m_A)}  \cdot \vcenter{\hbox{\includegraphics[width=1.6in]{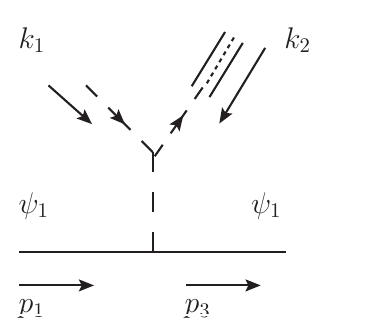}}}. \label{AITocI}
\end{eqnarray}
The last diagram originates from the corresponding terms in the (\ref{AToI}) where the pole of the middle propagator $\psi_2$ is canceled. Summing over such terms finally result in the third term of (\ref{AITocI}). The calculation processes are easy but lengthy, so we neglect the details here. We also point out that if we swap $\psi_1$ and $\psi_2$, (\ref{AITocI}) still holds.

From the above discussions, we can learn that through one direction of the differentiated propagator (\ref{dAdXi}), the vector or $I$ propagators successively transmute into ghosts and this will finally ends up to the other side of the (\ref{dAdXi}).  The processes look like we are ``tearing'' the vector-$I$ chain into a ghost chain plus another vector-$I$ chain  started with an $I$ while ended with a vector propagator. During this processes, terms within right-hand side of (\ref{NielsenRelying}) drop out. This can be sketched as
\begin{eqnarray}
\vcenter{\hbox{\includegraphics[width=1.5in]{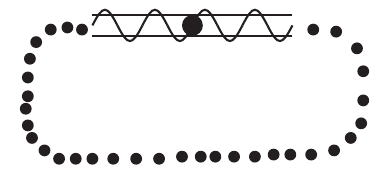}}} &\doteq&
-\vcenter{\hbox{\includegraphics[width=1.5in]{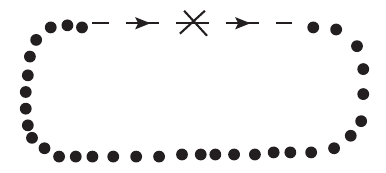}}} - (-i m_A)  \vcenter{\hbox{\includegraphics[width=1.5in]{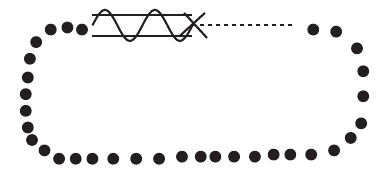}}} \nonumber \\
&+& (\text{Terms within right-hand side of (\ref{NielsenRelying})}) \label{AA_Tear}
\end{eqnarray}
where the dots indicate all combinations of the vector-$I$ chains, and we define
\begin{eqnarray}
\vcenter{\hbox{\includegraphics[width=2in]{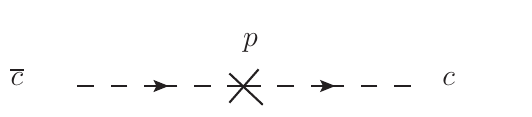}}} &=& \frac{i p^2}{\xi (p^2- \xi m_A^2)^2},  \label{cc_Modified} \\
\vcenter{\hbox{\includegraphics[width=2in]{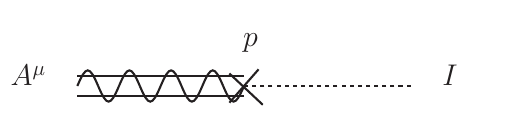}}} &=& \frac{i p^{\mu} }{(p^2- \xi m_A^2)^2}. \label{AI}
\end{eqnarray}
To understand (\ref{cc_Modified}) and (\ref{AI}), let us recollect and rewrite the (\ref{dAdXi})
\begin{eqnarray}
 \vcenter{\hbox{\includegraphics[width=2in]{PropagatorDADXi-eps-converted-to.pdf}}}  = \frac{-i p^{\mu} p^{\nu}}{(p^2- \xi m_A^2)^2}. \label{dAdXi_Repeated}
\end{eqnarray}
Notice that the $p^{\nu}$ will be absorbed into the ghost propagators in (\ref{cc_Modified}), and becomes the $(-i m_A)$ in the second term of (\ref{AA_Tear}), so it disappears in both (\ref{cc_Modified}) and (\ref{AI}).  The $p^{\mu}$ term remains in (\ref{AI}), however it will contract with the left part of the propagator in (\ref{cc_Modified}) to lead to the factor $\frac{p^2}{\xi}$. To see this, notice that (\ref{dAdXi}) can be reformulated into
\begin{eqnarray}
 \frac{-i p^{\mu} p^{\nu}}{(p^2- \xi m_A^2)^2} = (-i) \frac{ \xi p^{\mu} p^{\lambda} }{(p^2- \xi m_A^2)^2}  \frac{p_{\lambda}}{\xi p^2} p^{\nu}. \label{VVReformulated}
\end{eqnarray}
Notice that the vector propagator can be decomposed into
\begin{eqnarray}
& & \frac{-i}{p^2 - m_A^2} \left[ g_{\mu \nu} - \frac{p_{\mu} p_{\nu} (1-\xi) }{p^2 - \xi m_A^2} \right] \nonumber \\
&=& \frac{-i}{p^2 - m_A^2}  \left[ g_{\mu \nu} -  \frac{p_{\mu} p_{\nu} }{p^2}\right] + \frac{-i}{p^2 - m_A^2} \left[ \frac{p_{\mu} p_{\nu} }{p^2}  - \frac{p_{\mu} p_{\nu} (1-\xi)}{p^2 - \xi m_A^2} \right]  \nonumber \\
 &=& \frac{-i}{p^2 - m_A^2}  \left[ g_{\mu \nu} -  \frac{p_{\mu} p_{\nu} }{p^2}\right] + \frac{-i}{p^2 - \xi m_A^2} \left[ \frac{\xi p_{\mu} p_{\nu} }{p^2} \right]. \label{VPropReduced}
\end{eqnarray}
Since $p^{\mu} \cdot  \left[ g_{\mu \nu} -  \frac{p_{\mu} p_{\nu} }{p^2}\right]=0$, so during the prolonging processes,  like in (\ref{FirstStepR_Decompose}), or in (\ref{FirstStepRR_A})-(\ref{FirstStepRR_4Vertex}), etc.,  one can find out that if we only preserve the second term of (\ref{VPropReduced}), all the reduction processes remain unchanged. And this term coincide with the factor in (\ref{VVReformulated}), and will completely become a ghost propagator ended with the $p^{\lambda}$, contracting with the remained part of (\ref{VVReformulated}) to formulate the (\ref{cc_Modified}).

A subtle thing is that when $A^{\mu}$ is connected with an $R$-$I$ pair in (\ref{dAdXi_Repeated}), as in the third term of (\ref{FirstStepRR_ToBeCancelled}), the $p^2 = \xi m_A^2$ pole can be obliterated from the left side. We will argue that this will also be canceled when we manipulate the primitive ghost chains.

Now let us discuss about the primitive ghost chains. Ghost propagators should enclose into complete loops.  Each vector-$I$ loop is corresponding with two complete ghost loops with different directions.  The derivative of a ghost chain not only involves the derivatives of the propagators in (\ref{dPropgatorGhost}), but also includes the derivatives on vertices of (\ref{V_Rcc}), (\ref{V_RRcc}) and (\ref{V_IIcc}). Since each propagator can be paired up with one vertex, so we have, for example, when the ghost is emitting an $R$ just after the differentiated propagator, it is easy to calculate that
\begin{eqnarray}
& & \vcenter{\hbox{\includegraphics[width=1.2in]{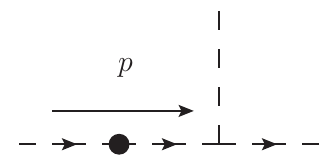}}} + \vcenter{\hbox{\includegraphics[width=0.9in]{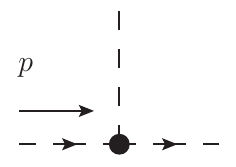}}} \nonumber \\
&=&  \vcenter{\hbox{\includegraphics[width=1.2in]{DGhostRGhost-eps-converted-to.pdf}}} + \frac{p^2-\xi m_A^2}{\xi(p^2 - \xi m_A^2)} \vcenter{\hbox{\includegraphics[width=0.9in]{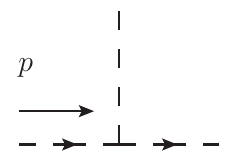}}} \nonumber \\
&=&\vcenter{\hbox{\includegraphics[width=1.2in]{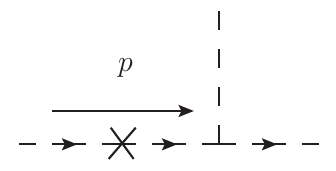}}}.
\end{eqnarray}
Similar results will be derived no matter the ghost propagator emits a single or double $R$, or double $I$, since all the corresponding vertices share the same structure proportional to $\xi$ (see (\ref{V_Rcc}), (\ref{V_RRcc}), (\ref{V_IIcc}), and their differentiated results (\ref{D_V_Rcc}), (\ref{D_V_RRcc}), and (\ref{D_V_IIcc}). 

Finally, differentiating a primitive ghost-chain gives the result of
\begin{eqnarray}
\vcenter{\hbox{\includegraphics[width=1.5in]{ccLoop_Modified-eps-converted-to.pdf}}}. \label{GhostChain_Canceled}
\end{eqnarray}
Since each complete vector-$I$ loop corresponds to two ghost chains towards two directions, so the first term of (\ref{AA_Tear}) cancels one of them.

The remained differentiated ghost chain can also be torn out into two chains. To illustrate this, let us recollect from (\ref{cIRI}) to (\ref{IIAA_C_Part}) that the half-ghost half-$I$ propagator of a ghost-chain can be moved backwards successively by summing over the following diagrams. We show the sketched processes below,
\begin{eqnarray}
& & \vcenter{\hbox{\includegraphics[width=4.2in]{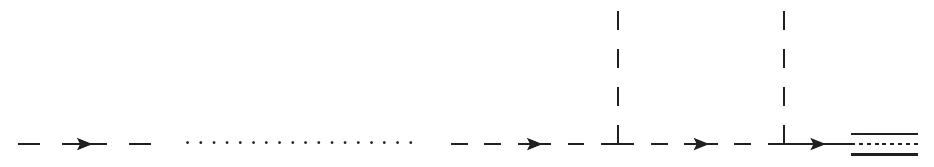}}} \nonumber \\
&+& \vcenter{\hbox{\includegraphics[width=4.2in]{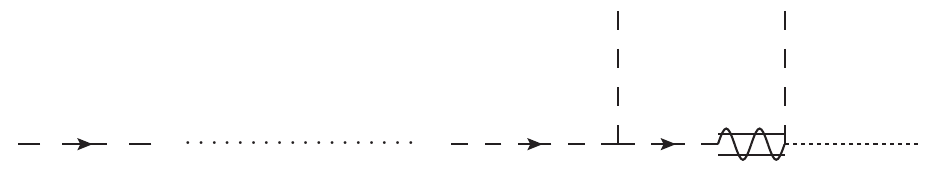}}} \nonumber \\
&\rightarrow&  \vcenter{\hbox{\includegraphics[width=4.2in]{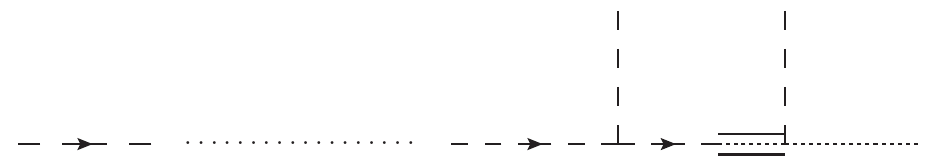}}}.  \label{Chain_Backwards}
\end{eqnarray}
The ``$\dots$'' here shadowed all kinds of emission of $R$'s and possible fermionic loops. This process can continue recursively,
\begin{eqnarray}
& & \vcenter{\hbox{\includegraphics[width=4.2in]{GhostChain_GI-eps-converted-to.pdf}}} \nonumber \\
&+& \vcenter{\hbox{\includegraphics[width=4.2in]{GhostChain_GAI-eps-converted-to.pdf}}} \nonumber \\
&+& \vcenter{\hbox{\includegraphics[width=4.2in]{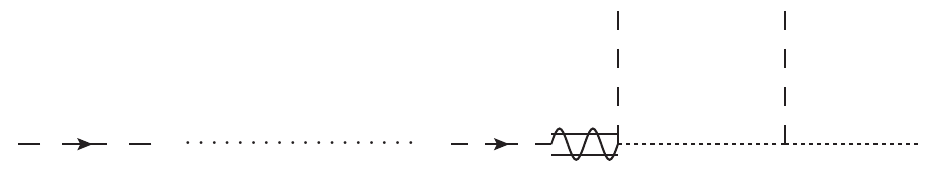}}}  \nonumber \\
&\rightarrow&  \vcenter{\hbox{\includegraphics[width=4.2in]{GhostChain_GII-eps-converted-to.pdf}}} \nonumber \\
&+& \vcenter{\hbox{\includegraphics[width=4.2in]{GhostChain_GAII-eps-converted-to.pdf}}} \nonumber \\
&\rightarrow& \vcenter{\hbox{\includegraphics[width=4.2in]{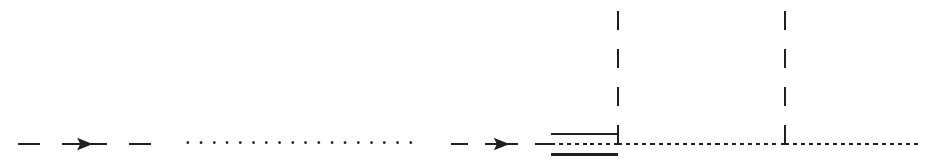}}},
\end{eqnarray}
etc., so
\begin{eqnarray}
& & \vcenter{\hbox{\includegraphics[width=4.2in]{GhostChain_GI-eps-converted-to.pdf}}} \nonumber \\
&+& \vcenter{\hbox{\includegraphics[width=4.2in]{GhostChain_GAI-eps-converted-to.pdf}}} \nonumber \\
&+& \vcenter{\hbox{\includegraphics[width=4.2in]{GhostChain_GAII-eps-converted-to.pdf}}}  \nonumber \\
&+& \dots \nonumber \\
&\rightarrow& \vcenter{\hbox{\includegraphics[width=4.2in]{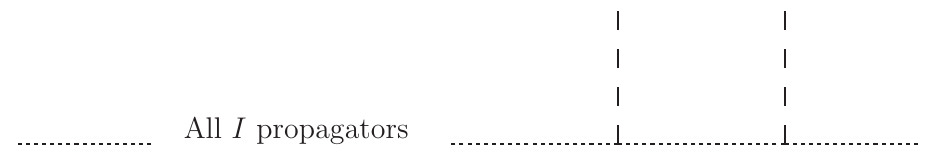}}} \cdot (-i m_A). \label{VectorLadders}
\end{eqnarray}

Notice that according to (\ref{ABeginAFinalReduction}), 
\begin{eqnarray}
&+& \vcenter{\hbox{\includegraphics[width=4.2in]{GhostChain_GAI-eps-converted-to.pdf}}} \nonumber \\
&+& \vcenter{\hbox{\includegraphics[width=4.2in]{GhostChain_GAII-eps-converted-to.pdf}}}  \nonumber \\
&+& \dots \nonumber \\
&\rightarrow& \vcenter{\hbox{\includegraphics[width=3.7in]{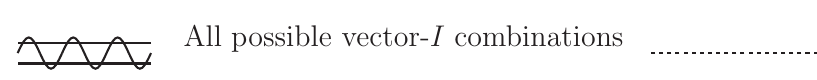}}} \nonumber \\
&-& \vcenter{\hbox{\includegraphics[width=3.7in]{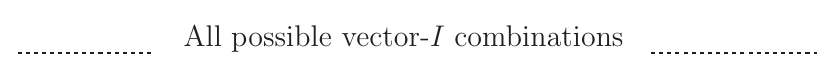}}} \cdot (-i m_A) \nonumber \\
&+& \vcenter{\hbox{\includegraphics[width=3.7in]{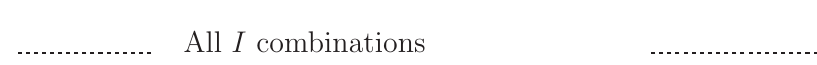}}} \cdot (-i m_A).
\end{eqnarray}
Substitute this into (\ref{VectorLadders}) and move all the terms other than the ghost chain to the right-hand side, we have,
\begin{eqnarray}
& & \vcenter{\hbox{\includegraphics[width=3.7in]{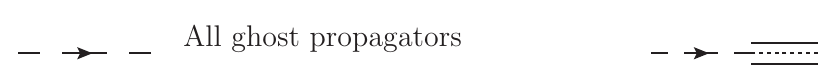}}} \nonumber \\
&\rightarrow& -  \vcenter{\hbox{\includegraphics[width=3.7in]{GhostChain_AStart-eps-converted-to.pdf}}} \nonumber \\
&+& \vcenter{\hbox{\includegraphics[width=3.7in]{GhostChain_IStart-eps-converted-to.pdf}}} \cdot (-i m_A). \label{GhostChainTeared}
\end{eqnarray}
Curve the above processes into diagram loops, and notice that orientation of the remained ghost loop should be opposite with the (\ref{GhostChain_Canceled}),  and carefully manipulate with the coefficients, we have
\begin{eqnarray}
\vcenter{\hbox{\includegraphics[width=1.5in]{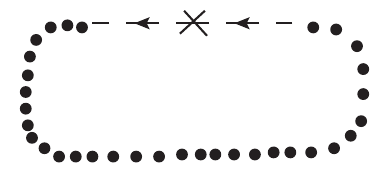}}} &\doteq& (-i m_A)  \vcenter{\hbox{\includegraphics[width=1.5in]{AILoop-eps-converted-to.pdf}}} - \vcenter{\hbox{\includegraphics[width=1.5in]{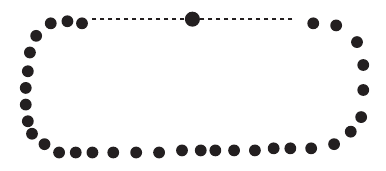}}} \nonumber \\
&+& (\text{Terms within right-hand side of (\ref{NielsenRelying})}) \label{CC_Tear}
\end{eqnarray}
The first term in (\ref{CC_Tear}) cancels with the second term in (\ref{AA_Tear}), and the second term in (\ref{CC_Tear}) cancels with the differentiated $I$ loop induced by (\ref{dPropgatorI}).  Here the terms separating the bulk and the $C$-parts also appears in (\ref{CC_Tear}) due to the similar reason in (\ref{AA_Tear}), although we omitted them during the deriving processes (\ref{Chain_Backwards})-(\ref{GhostChainTeared}).

Before we get the conclusion, let us patch two leaks when we acquire the (\ref{GhostChainTeared}). At the right-end of the diagrams, we might encounter patterns like (\ref{OneUnitBackward}). However (\ref{OneUnitBackward}) omitted the term to kill the $k_n^2 = \xi m_A^2$ pole, leaving us a term which will be canceled by the previous term that we have mentioned shortly after (\ref{VPropReduced}) that the vector boson's pole at the  $A^{\mu}$ side in (\ref{dAdXi_Repeated}) is obliterated. We neglect the detailed calculation of this leak, and emphasize another one which is more important.

At the left-end of the diagrams in (\ref{GhostChainTeared}), we sometimes have to calculate
\begin{eqnarray}
\vcenter{\hbox{\includegraphics[width=2.2in]{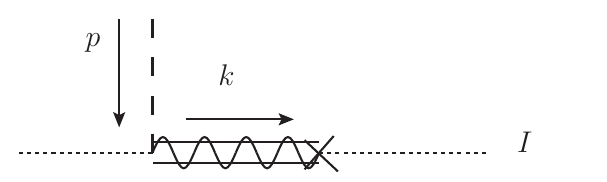}}}.
\end{eqnarray}
We neglect the details and only point out that the vertex induces a $2 p \cdot k = (p+k)^2 - p^2 - k^2$, generating a $k^2-\xi m_A^2$ to kill the vector propagator, shrinking it into a point, and finally resulting a differentiated $I$-$I$-$R$ vertex,
\begin{eqnarray}
\vcenter{\hbox{\includegraphics[width=2.2in]{IR_PropagatorAI-eps-converted-to.pdf}}} \rightarrow - \vcenter{\hbox{\includegraphics[width=1.7in]{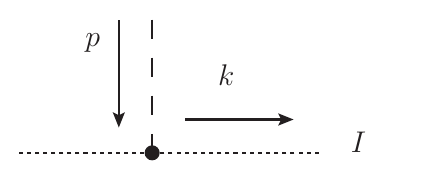}}}. \label{IIR_Diff_Canceled}
\end{eqnarray}
This term is canceled by the vector-$I$ loop induced by the differentiated $I$-$I$-$R$ vertex in (\ref{D_V_RII}).

Another subtle diagram is induced by both the beginning and the end of the (\ref{ABeginAFinalReduction}). (\ref{dAdXi}) might encounter connections like the following pattern,
\begin{eqnarray}
\vcenter{\hbox{\includegraphics[width=2.4in]{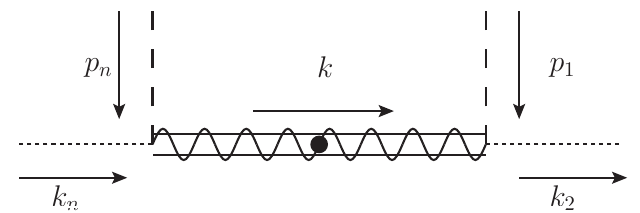}}}.
\end{eqnarray}
Again, the $p_n \cdot k$ and $p_1 \cdot k$ can induce terms countering the $k$ propagators, which means
\begin{eqnarray}
\vcenter{\hbox{\includegraphics[width=2.4in]{IR_PropagatorDADXi_IR-eps-converted-to.pdf}}} \rightarrow-\vcenter{\hbox{\includegraphics[width=1.4in]{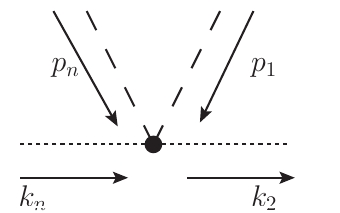}}} . \label{IIRR_Diff_Canceled}
\end{eqnarray}
This term is also canceled by the vector-$I$ loop induced by the differentiated $I$-$I$-$R$-$R$ vertex in (\ref{D_V_RIII}).

Now we found that all the terms that are irrelevant to the right-hand side of (\ref{NielsenRelying}) have been canceled by each other to disappear. One can compare and find out that the terms in  (\ref{AA_Tear}) give rise to the $\langle I(x) c(x) \overline{c}(y) (\partial_{\mu} A^{\mu}) \rangle_{\text{1PI}}$ terms, (\ref{CC_Tear}) results in the $\langle I(x) c(x) \overline{c}(y)  g \xi v I \rangle_{\text{1PI}}$ terms, and the third term of (\ref{FirstStepRR_ToBeCancelled}), as we have mentioned there, contributes to the $\langle I(x) c(x) \overline{c}(y) (g \xi R I) \rangle_{\text{1PI}}$ terms. One can compare the coefficients with the expanded couplings at the right-hand side of (\ref{NielsenRelying}), and might find out that our calculations of the factors are twice as the (\ref{NielsenRelying}) predicts. This is because we only considered one of the two orientations of the (\ref{dAdXi}) to prolong our ghost chain. Actually both directions are possible, so averaging both these orientations supplements the factor of $\frac{1}{2}$ that we need.

\section{Generalization to $R_{\xi}$ gauges, non-abelian groups, finite temperature cases, multiple Higgs  and fermionic multiplets}

In the previous sections, we relied on the simple abelian toy model. However in a more practical situation, we have to get involved with the non-abelian gauge group. We might also have to count in the thermal effects if the processes we want to evaluate happen in the early universe. A bunch of Higgs bosons or fermions with complicated mixing patterns might be encountered. $R_{\xi}$ gauges might also be the case because it is more familiar. Our previous discussions can be generalized to all these cases. We are going to discuss them separately, although we might not show the detailed proof.

All the previous discussions can be directly generalized to the finite temperature cases when one performs the effective potential evaluations. With the imaginary time formalism,  the metric shifts from the Minkowski form to the Euclidean form, so only the inner products have to be updated. The space-time integral also has to be changed from $\int d^4 p$ into $\sum\limits_i \int d^3 \vec{p}$,  where $\sum\limits_i$ means to sum over all the Matsubara frequencies\cite{Laine:2016hma, Bellac:2011kqa, Quiros:1999jp}. Obviously all our discussions are inside the integral or summation symbols, so all our results remain intact in this case.

For the multiple Higgs or fermion cases when complicated mixings arise, one has to calculate on the mass-eigenstate basis, and adjoin all the corresponding couplings with proper mixing matrix elements.  For all the previous diagrams with scalar or fermionic inner lines, one has to sum over all possible propagators with different mass eigen-states. Although we neglect the detailed evaluations, we just state that all the previous discussions are still valid by applying the unitarity property of the mixing matrices.

For the $R_{\xi}$ gauge and non-abelian situations, things are a little more complicated, so we address the details in the two following subsections.

\subsection{The $R_{\xi}$ gauge discussions}

All the above discussions can be simply cast into the $R_{\xi}$ gauges.  Although we do not repeat the whole processes in the $R_{\xi}$ gauge, we just list the key differences between the two kinds of gauges. Compared with the $\overline{R}_{\xi}$ gauges (\ref{GaugeFixing}),  the gauge fixing term does not include $R$,
\begin{eqnarray}
\mathcal{L}_{\text{g.f., }R_{\xi}} &=& -\frac{1}{2 \xi} F_{R_{\xi}}^2, \label{GaugeFixing_NoBar} \\
F_{R_{\xi}} &=& \partial_{\mu} A^{\mu} - g \xi v I.
\end{eqnarray}
Therefore, (\ref{V_ARI}) recovers to $g (p_1^{\mu} - p_2^{\mu})$,  and the $\xi$-dependent terms in (\ref{RRII_Vertex}) and (\ref{RII_Vertex}) disappear. The ghost terms become
\begin{eqnarray}
\mathcal{L}_{\text{f.p., }R_{\xi}} = - \overline{c} [ \square + \frac{1}{2} \xi g^2 R^2 ] c, \label{GhostInteraction_Rxi}
\end{eqnarray}
so that (\ref{V_RRcc}), (\ref{V_IIcc}) disappears, and (\ref{V_Rcc}) waives the factor of $2$ to become $-i g^2 \xi$.

The manipulation of the vector-$R$-$I$ vertices might be a little bit tricky. For example,  in (\ref{AFirstStepII_t}), the $2 g k_1 \cdot (-k_1 - p_1) = g [k_1^2 - (k_1+p_1)^2 - p_1^2]$ should be replaced with 
\begin{eqnarray}
g k_1 \cdot (-k_1 - 2 p_1) = g [ p_1^2 - (k_1+p_1)^2 ].
\end{eqnarray}
However, in this case only the poles of the $p_1$ and $k_1+p_1$ propagators will be obliterated. Therefore $\langle g \xi RI \rangle_{\text{1PI}}$, as well as the processes of (\ref{IIR_Diff_Canceled}), (\ref{IIRR_Diff_Canceled}) are unnecessary, just as the expected Nielsen identity within the frameset of the $R_{\xi}$ gauges
\begin{eqnarray}
\xi \frac{\partial \Gamma_{R_\xi} [R, \xi]}{\partial \xi} = -\int d^4 x \frac{\delta \Gamma_{R_{\xi}}}{\delta R(x)} C_{R, R_{\xi}} (x), \label{NielsenRelying_Rxi}
\end{eqnarray}
where
\begin{eqnarray}
C_{R, R_{\xi}}(x)  &=& -\frac{i}{2} \int d^4 y \langle I(x) c(x) \overline{c}(y) (\partial_{\mu} A^{\mu} + g \xi v I) \rangle_{\text{1PI}}. \label{CPartOfNielsen_Rxi}
\end{eqnarray}

\subsection{Generalization to the non-abelian gauge groups}

The most prominent difference between nthe on-abelian gauged group models with the abelian ones are the self-interactions among the gauge bosons and the additional interactions between the ghosts and the gauge bosons. These accumulate the complexities of the proof, and in this paper, we only express the main algorithm without anatomizing every details as before.

The peculiar vertices of the non-abelian gauge interactions are listed below,
\begin{eqnarray}
\vcenter{\hbox{\includegraphics[width=1.4in]{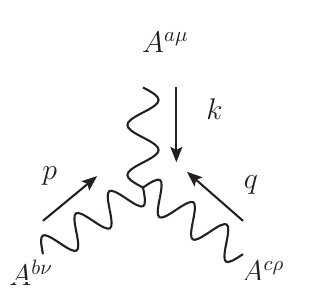}}} &=& g f^{abc} [g^{\mu \nu} (k-p)^{\rho} + g^{\nu \rho} (p-q)^{\mu} + g^{\rho \mu} (q-k)^{\nu}],   \label{V_AAA}\\
\vcenter{\hbox{\includegraphics[width=1.4in]{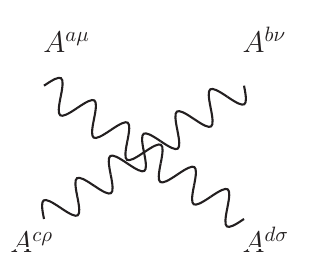}}} &=& \begin{array}{c}
-i g^2 [f^{abe} f^{cde} (g^{\mu \rho} g^{\nu \sigma} - g^{\mu \sigma} g^{\nu \rho}) \\
+ f^{ace} f^{bde} (g^{\mu \nu} g^{\rho \sigma} - g^{\mu \sigma} g^{\nu \rho}) \\
+ f^{ade} f^{bce} (g^{\mu \nu} g^{\rho \sigma} - g^{\mu \rho} g^{\nu \sigma}) ]
\end{array}, \label{V_AAAA}\\
\vcenter{\hbox{\includegraphics[width=1.4in]{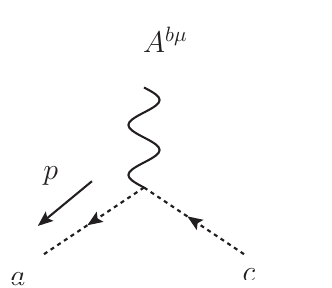}}} &=& -g f^{abc} p^{\mu},
\end{eqnarray}
where $a,b,c$ are the group indices, and $f^{abc}$ are the structure constants. 

During the prolonging processes of the ghost chain, if one encounters (\ref{V_AAA}) or (\ref{V_AAAA}), he should select  a route to prolong the chain.  Cases are that some particular terms are selected for a particular route, while others are for other possible routes. For example, if one encounters
\begin{eqnarray}
\vcenter{\hbox{\includegraphics[width=1.4in]{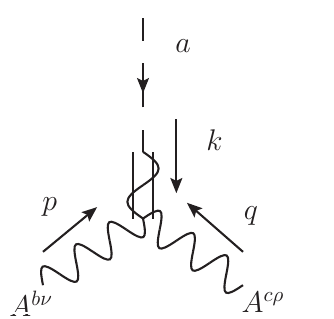}}} &=& g f^{abc} [k^{\rho} (q+k)^{\nu} + k^{\nu} (-k-p)^{\rho} + g^{\nu \rho} k \cdot (p-q)],
\end{eqnarray}
Where we have applied the trick to add $2 k^{\rho} k^{\nu}$ in the first term, while deduct it in the second term. Notice that $k \cdot (p-q) = q^2-p^2 = p^2 - m_A^2 + q^2 - m_A^2$, killing the two poles of the $A^{b \nu}$ and $A^{c \rho}$ propagators separately, and these terms will further be canceled by other diagrams. We only concern the remained $g f^{abc} [k^{\rho} (q+k)^{\nu} + k^{\nu} (-k-p)^{\rho}]$ term. If one chooses $A^{b \nu}$ as the ghost chain to prolong, $g f^{abc} [k^{\rho} (q+k)^{\nu}]$ should be selected, while the term of $g f^{abc} [ k^{\nu} (-k-p)^{\rho}]$  is for the alternative route. For example, if we select $A^{b \nu}$, one can finally acquire the vertex
\begin{eqnarray}
\vcenter{\hbox{\includegraphics[width=1.4in]{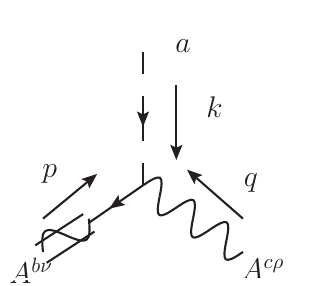}}}.
\end{eqnarray}

For the (\ref{V_AAAA}),  for example
\begin{eqnarray}
\vcenter{\hbox{\includegraphics[width=1.4in]{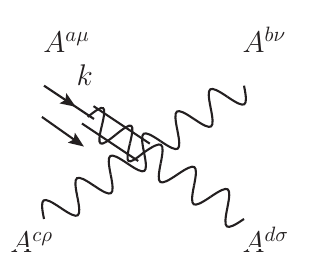}}} &=& \begin{array}{c}
-i g^2 [ k^{\rho} (f^{abe} f^{cde}-f^{a d e} f^{b c e}) \\
+ k^{\nu} ( f^{ace} f^{bde} + f^{a d e} f^{b c e}) \\
+ k^{\sigma} (-f^{a b e} f^{c d e} - f^{a c e} f^{b d e}) ].
\end{array}
\end{eqnarray}
If, e.g., one selects the $A^{c \rho}$ route to prolong the ghost chain, then the first term should be adopted. The remained $A^{b \nu}$ and $A^{d \sigma}$ are then treated as two ``scalars'',  so the tricks from (\ref{FirstStepRR_Cs}) to (\ref{FirstStepRR_4VertexResult}) can be applied with their vector-$I$-$R$ vertices replaced with (\ref{V_AAA}).

\section{Suggestions about improvements of the super-daisy diagram resummation method}

The effective potential is equivalent to the $\Gamma$-function,al which is critical in calculating various observables in which the concept of the ``particle'' becomes vague. For example, the tunneling rates between two vacuums, and the gravitational waves generated through these processes.  The inflation of the universe might also be driven by some scalars rolling slowly along their effective potentials. Conventional evaluations involve calculating the Colemann-Weinberg potential, in which all one-loop diagrams are resummed. 

\begin{figure}
\begin{equation}
\vcenter{\hbox{\includegraphics[width=0.4\textwidth]{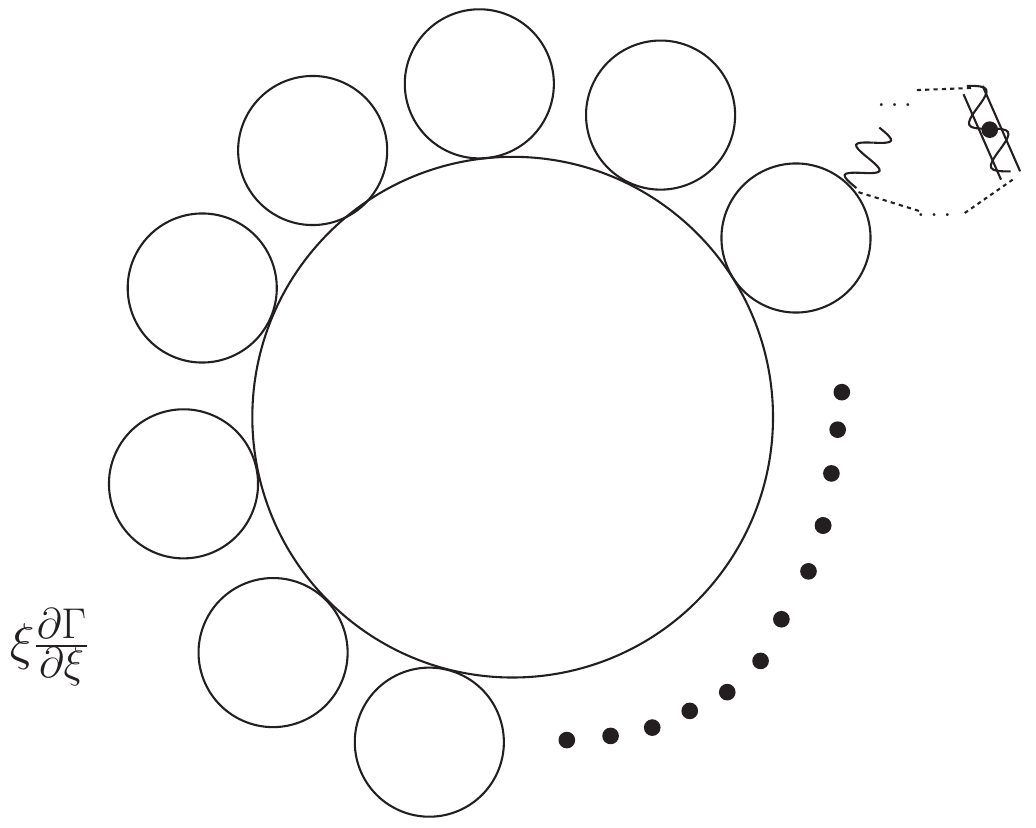}}} \rightarrow
\vcenter{\hbox{\includegraphics[width=0.5\textwidth]{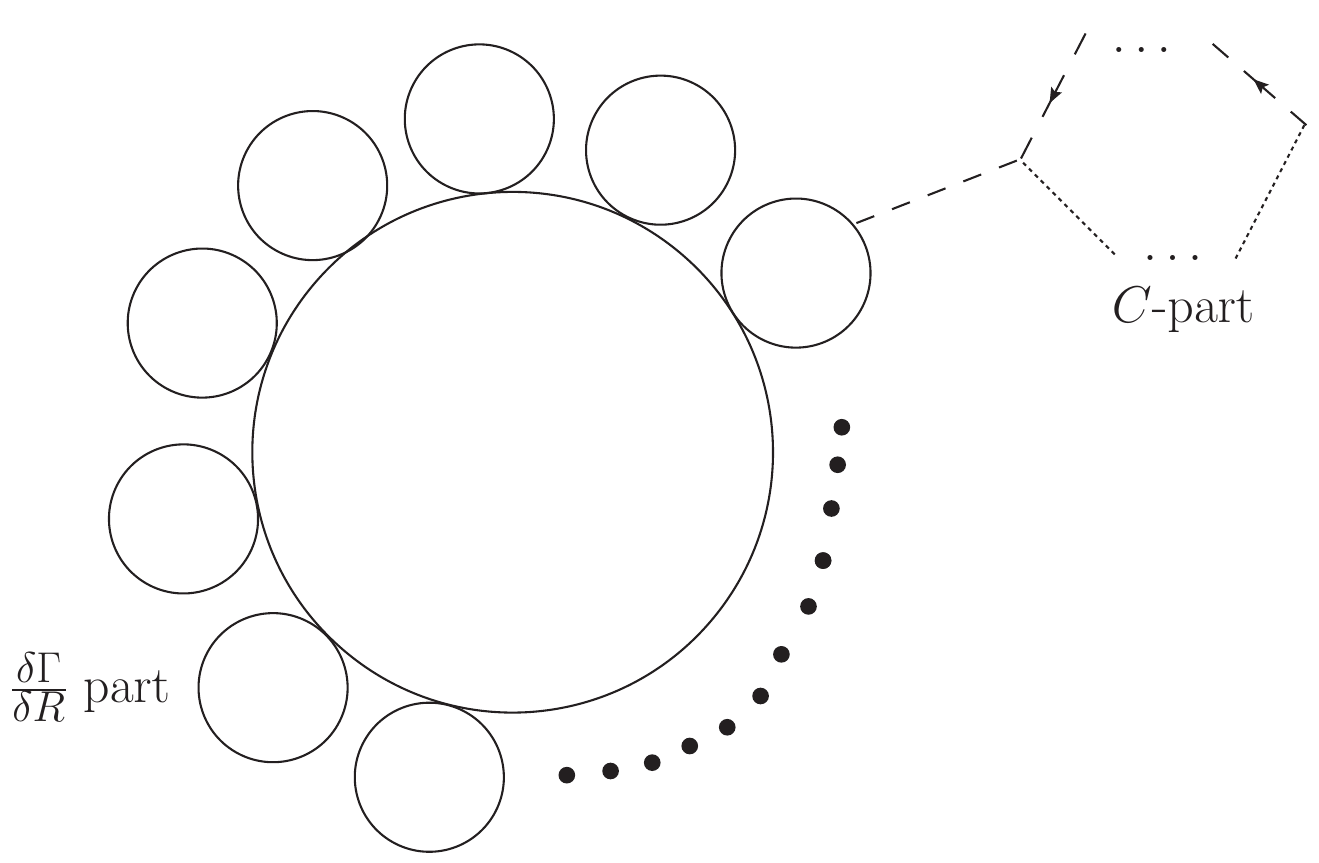}}} \nonumber
\end{equation}
\caption{Sketched diagrams explaining why one layer of daisy resummation is insufficient to fit the (\ref{NielsenRelying}) rigorously.  The right panel includes only one layer of daisy diagrams, however through clipped with a $C$-part diagram, it will result in an additional layer of ringlets in the $\xi \frac{\partial \Gamma}{\partial \xi}$ components.} \label{Daisy_Onelayer}
\end{figure}

Sometimes daisy diagrams are considered to improve the Colemann-Weinberg potential.  However it is easy to learn from our previous discussions that only one layer of daisy ringlets is insufficient for an effective potential to satisfy (\ref{NielsenRelying}) rigorously. If, for example, $\Gamma_{\text{daisy}}$ is the effective potential in which all daisy diagrams have been included, and if one of the daisy ringlets is connected with an external line of $R$,  then the $\xi \frac{\partial \Gamma}{\partial \xi}$ of the left-hand side of (\ref{NielsenRelying}) is inevitably clipped with an additional loop through the $C$-part on the right-hand side. Just as illustrated in Fig.~\ref{Daisy_Onelayer}. Therefore, in the literature, people expand $\Gamma$ into series of $\hbar^n$, where $n$ indicates the number of loops. Any finite expansions on $\hbar$ orders never rigorously satisfy the (\ref{NielsenRelying}), since the orders of $\hbar$ never balance on both sides of it, so some remained terms arise. Therefore people usually drop out the higher order terms, and reach a concessional result which satisfies the Nielsen identity ``order by order''.

\begin{figure}
\includegraphics[width=2.7in]{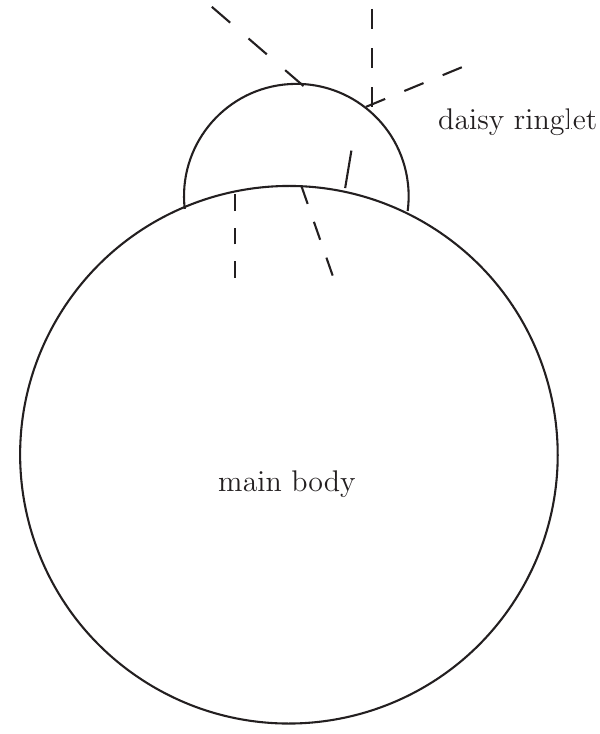}
\caption{An example that a daisy ringlet shares the multiple common propagators with the main body}
\label{Daisy_MCS}
\end{figure}

To acquire an effective potential satisfying the Nielsen identity up to all $\hbar$ orders, we have to resum  at least the diagrams with all possible connections of the ringlets, so called the ``super-daisy'' diagrams.  We can evaluate each ringlet up to a fixed $\hbar$ order, and then stack them to form cactus-shaped objects. In the literature, ``daisy-diagram'' or ``daisy resummation'' sometimes indicates that the ringlets might share multiple common internal propagators with the main body, like in Fig.~\ref{Daisy_MCS}. In this paper, we focus on the case that each daisy ringlet shares only one common propagator or only one common vertex with other parts of the diagram, either the main body, or another ringlet. This is exactly the structure of the ``waist'' of a ``gourd'' that we have described in Fig.~\ref{GourdSketch}. We might call these a particular sort of super-daisy diagrams.

Since now, a complete resummation of all possible diagrams are far beyond human being's capability. People have to abandon an infinite number of diagrams. Therefore, practically each ringlet is calculated up to a particular order of a parameter, either the coupling constants or the $\hbar$. Then we stack these ringlets into super-daisy diagrams with the waist structures.

One might naively believe that if we resum only these particular sort super-daisy diagrams that we have discussed just now, we can acquire an effective potential satisfying (\ref{NielsenRelying}), since if we are prolonging the ghost-chains, the $C$-part and the bulk part separates when the chain hits each waist. However, this idealism is broken by the fact that we need additional diagrams to cancel some exotic terms. For example,
\begin{eqnarray}
\vcenter{\hbox{\includegraphics[width=2.0in]{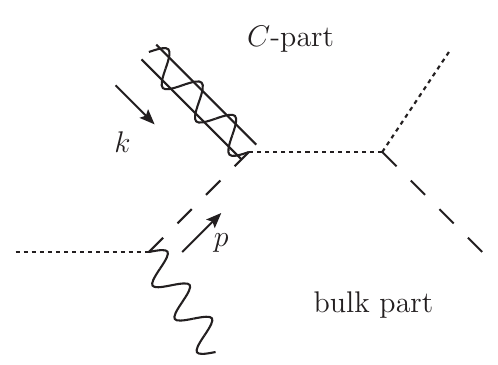}}}  \supset 2 g k \cdot p = (k+p)^2 - k^2 - p^2. \label{SketchedWaist}
\end{eqnarray}
Again, this $p^2$ will generate a $p^2-m_R^2$ factor to cancel the $p$-propagator. Sometimes this requires inevitably such a diagram like
\begin{eqnarray}
\vcenter{\hbox{\includegraphics[width=2.0in]{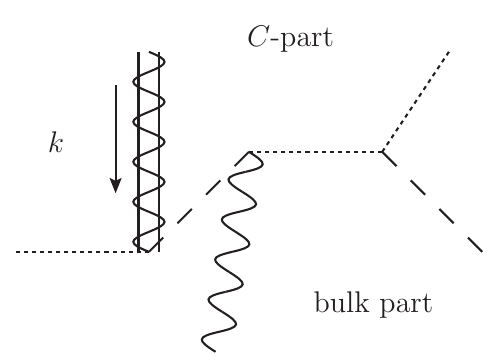}}} \label{SketchedWaist_Nongourd}
\end{eqnarray}
to cancel it.  However,  (\ref{SketchedWaist_Nongourd}) breaks the gourd waist structure since the common lines between two ringlets take two propagators. Therefore, only a resummation of our particular sort of diagrams will never fit the (\ref{NielsenRelying}) rigorously.

To cure this problem,  one might think of directly getting rid of the irrelevant $p^2-m_R^2$ term. However,  before the differentiating operation, there will be no double-lined half-vector propagator (\ref{HalfA}) appeared in the diagram, making it difficult to separate the exotic terms. Fortunately, as we have mentioned, if we rewrite the composition of the gauge boson's propagator from (\ref{VPropReduced}) here,
\begin{eqnarray}
& & \frac{-i}{k^2 - m_A^2} \left[ g_{\mu \nu} - \frac{k_{\mu} k_{\nu} (1-\xi) }{k^2 - \xi m_A^2} \right] \nonumber \\
 &=& \frac{-i}{k^2 - m_A^2}  \left[ g_{\mu \nu} -  \frac{k_{\mu} k_{\nu} }{k^2}\right] + \frac{-i}{k^2 - \xi m_A^2} \left[ \frac{\xi k_{\mu} k_{\nu} }{k^2} \right], \label{VPropReduced_Rewrite}
\end{eqnarray}
we know that it is the $\xi$-dependent time-like $\left[ \frac{\xi k_{\mu} k_{\nu} }{k^2} \right]$ term that actually works in all the above discussions since $k^{\mu}   \left[ g_{\mu \nu} -  \frac{k_{\mu} k_{\nu} }{k^2}\right]  \equiv 0$. Therefore, if we separate the propagator into the ``Landau part'' and the ``timelike part'' as we call them,
\begin{eqnarray}
& & \vcenter{\hbox{\includegraphics[width=2in]{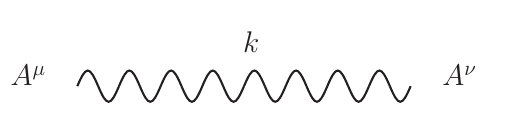}}} \nonumber \\
&=& \vcenter{\hbox{\includegraphics[width=2in]{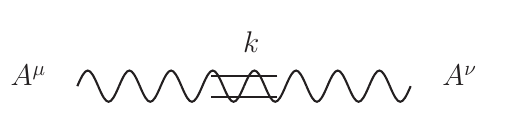}}}  + \vcenter{\hbox{\includegraphics[width=2in]{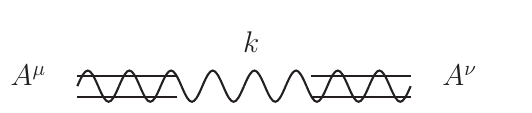}}}, \label{VProp_Decomposed}
\end{eqnarray}
where
\begin{eqnarray}
 \vcenter{\hbox{\includegraphics[width=2in]{PropagatorA_Landau-eps-converted-to.pdf}}} = \frac{-i}{k^2 - m_A^2}  \left[ g_{\mu \nu} -  \frac{k_{\mu} k_{\nu} }{k^2}\right],
\end{eqnarray}
is the Landau part, and the 
\begin{eqnarray}
 \vcenter{\hbox{\includegraphics[width=2in]{PropagatorA_Timelike-eps-converted-to.pdf}}} =  \frac{-i}{k^2 - \xi m_A^2} \left[ \frac{\xi k_{\mu} k_{\nu} }{k^2} \right],
\end{eqnarray}
is the timelike part, then before differentiating, (\ref{SketchedWaist}) can be reduced into
\begin{eqnarray}
& & \vcenter{\hbox{\includegraphics[width=2.0in]{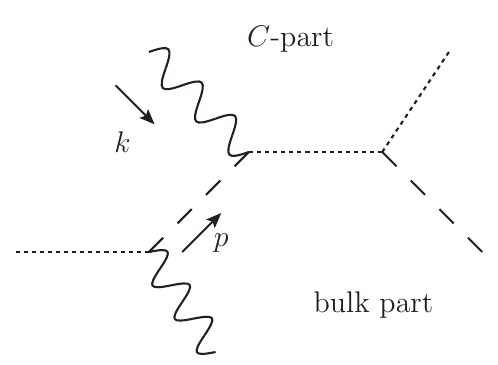}}}  \nonumber \\
&=& \vcenter{\hbox{\includegraphics[width=2.0in]{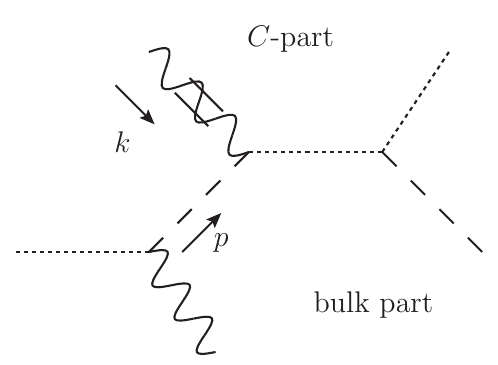}}} + \vcenter{\hbox{\includegraphics[width=2.0in]{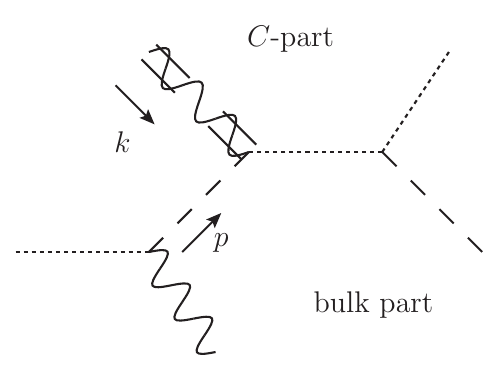}}}. \label{SketchedWaist_Reduced}
\end{eqnarray}
The Landau part can be calculated separately but it will not contribute to the $C$-part of the diagrams after differentiating. It is sometime tolerable to get rid of this term to sacrifice some precision. However, the vector propagator in the second timelike term can contribute to a $2 g k \cdot p$, and again takes the chance to cancel the $p^2 = m_R^2$ pole. We can then safely drop out this term so that after differentiating, such term also vanish. Thus the waist-structure breaking diagrams in (\ref{SketchedWaist_Nongourd}) are no longer essential.

Generally, all vector bosons connected to the waist of the gourd can be decomposed into the Landau part and the timelike part as described in (\ref{VProp_Decomposed}). The timelike part may then contribute a momentum contracting with the following momentum. This can generate terms killing each of the poles of the corresponding propagator connecting to this vector boson to give three terms respectively, and one can just get rid of the irrelevant term to avoid the appearance of a waist-structure breaking diagram.

When the waist is the fermionic type,  e.g., like in (\ref{AITocI}), we still have the chance to kill the pole of the fermionic propagator. For example, 
\begin{eqnarray}
  \vcenter{\hbox{\includegraphics[width=1.8in]{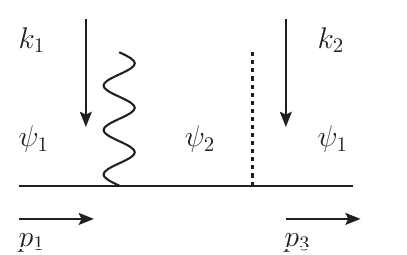}}}  =   \vcenter{\hbox{\includegraphics[width=1.8in]{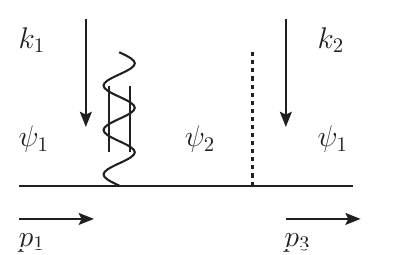}}}   +   \vcenter{\hbox{\includegraphics[width=1.8in]{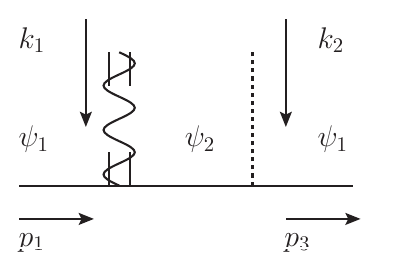}}},
\end{eqnarray}
and the second timelike term contributes to a
\begin{eqnarray}
& & \frac{i(p\!\!\!/_1+m_1)}{p_1^2-m_1^2} k\!\!\!/_1 \frac{i(p\!\!\!/_1+k\!\!\!/_1+m_2)}{(p_1+k_1)^2-m_2^2} 
\nonumber \\
&=& \frac{i(p\!\!\!/_1+m_1)}{p_1^2-m_1^2} [(p\!\!\!/_1 + k\!\!\!/_1 + m_2) -  (p\!\!\!/_1+m_1) + (m_1-m_2)] \frac{i(p\!\!\!/_1+k\!\!\!/_1+m_2)}{(p_1+k_1)^2-m_2^2}.
\end{eqnarray}
The  second term in the middle squared bracket cancels the $p^2-m_1^2$ pole, which requires another waist-structure breaking diagram to kill it. However, as we have addressed before, directly dropping out this term does not break our previous discussions about the Nielsen identity, and the appearance of the waist-breaking diagram becomes unnecessary.

For the primitive ghost loop, things are a little bit subtle.  There is no apparent vector propagator to formulate momentum inner products. However,  Both (\ref{ABeginAFinalReduction}) and (\ref{GhostChainTeared}) allow us to anatomize a confined area of a primitive ghost chain, without disturbing the other parts. Since half of the contribution from the primitive ghost chain is cancelled by (\ref{ABeginAFinalReduction}), and the other half is reduced by (\ref{GhostChainTeared}), we have to separate the primitive ghost chain into two equal halves, and utilize (\ref{ABeginAFinalReduction}) and (\ref{GhostChainTeared}) to anatomize each of the chain. We can do this surgery around the waist. For example, the half exerted by the (\ref{GhostChainTeared}) can be reduced to
\begin{eqnarray}
& & \vcenter{\hbox{\includegraphics[width=2.0in]{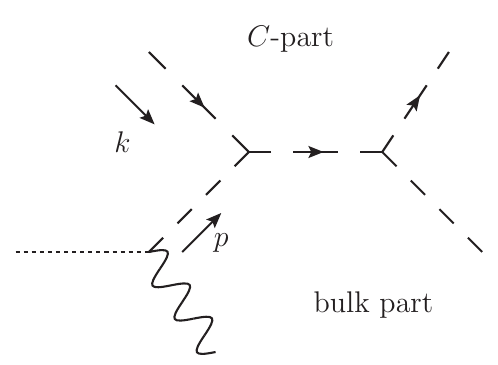}}} +  \vcenter{\hbox{\includegraphics[width=2.0in]{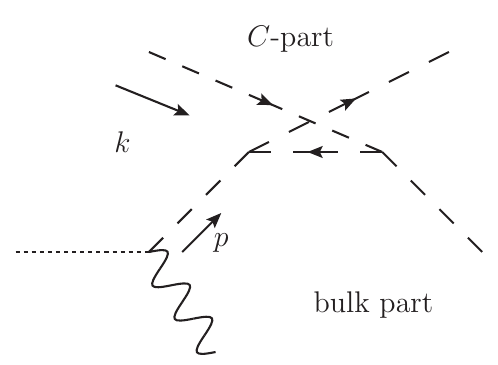}}} \nonumber \\
&+& \vcenter{\hbox{\includegraphics[width=1.5in]{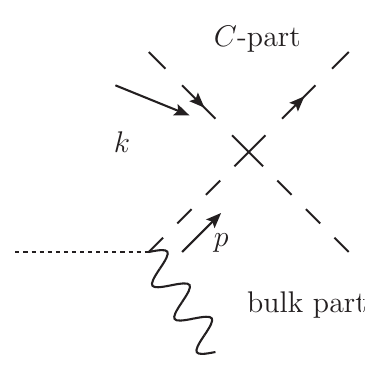}}} \nonumber \\
&\rightarrow&   \vcenter{\hbox{\includegraphics[width=2.0in]{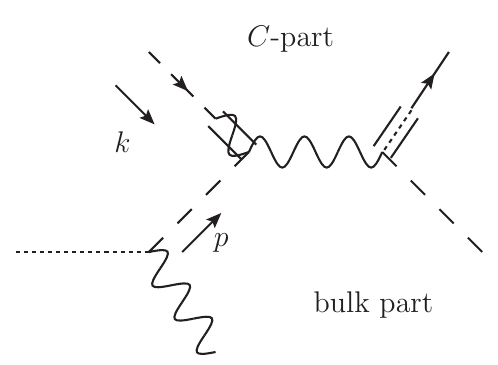}}} + \vcenter{\hbox{\includegraphics[width=2.0in]{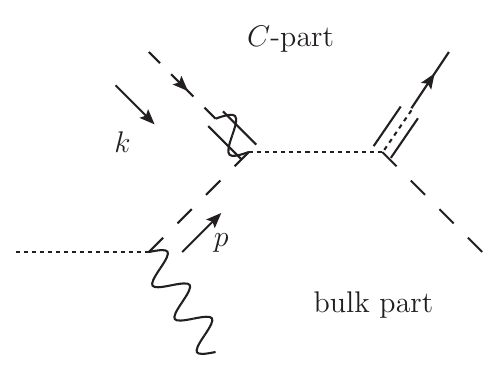}}} \nonumber \\
&+&  \vcenter{\hbox{\includegraphics[width=2.0in]{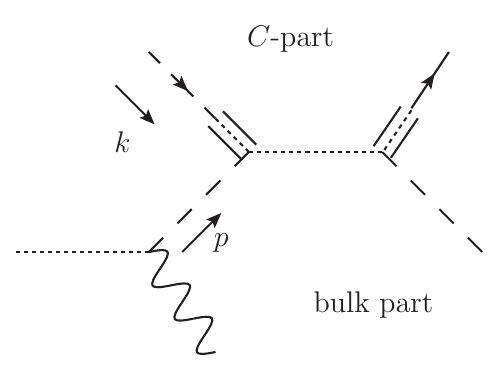}}} + \dots. \label{SketchedWaist_Ghost_Reduced}
\end{eqnarray}
The above expression is only a sketched discussion. One has to supplement the coefficients and the neglected diagrams during practical evaluations.  However, it is clear that some diagrams induce cancellations to the irrelevant propagator poles. For example,
\begin{eqnarray}
 \vcenter{\hbox{\includegraphics[width=2.0in]{SketchedWaist_ghostVVI-eps-converted-to.pdf}}}  \supset 2 g k \cdot p = (k+p)^2 - k^2 - p^2. \label{Sketched_Ghost}
\end{eqnarray}
again takes the chance to cancel the $p^2 = m_R^2$ pole, and can be dropped out to avoid the involvement of the waist-structure breaking diagrams.

The ghosts themselves might also form a waist connected with an $R$-propagator,
\begin{eqnarray}
& & \vcenter{\hbox{\includegraphics[width=1.8in]{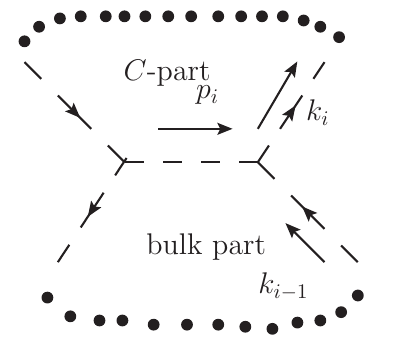}}} \rightarrow \vcenter{\hbox{\includegraphics[width=1.8in]{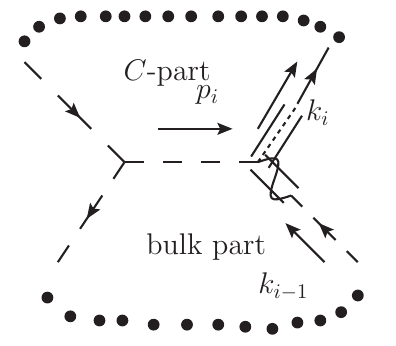}}} + \vcenter{\hbox{\includegraphics[width=1.8in]{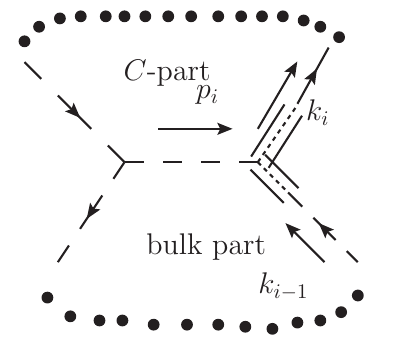}}}  \nonumber \\
&+& \dots.
\end{eqnarray}
One can also anatomize the ghost propagators near the waist, and a vector propagator appears to kill some of the closest poles.  Terms involving the waist-structure breaking diagrams are also needed to be dropped out

Therefore, the diagrammatic study of the Nielsen identity tells us that a super-daisy resummation with some particular manipulations near the waists might help us acquire an effective potential satisfying the Nielsen identity up to all $\hbar$ orders.

\section{Summary and future prospect}

In this paper, we have relied on a gauge $U(1)$ toy model to prove the Nielsen identity diagrammatically rather than the path integral method usually appeared in the literature. From the proof one can realize clearly how the partially $\xi$-differentiated 1PI diagrams at the left-hand side of (\ref{NielsenRelying}) divides into two parts to formulate the right-hand side of (\ref{NielsenRelying}) respectively. The conversion from the vector-$I$ chains into the ghost chain and the cancellation of the other exotic terms  among diagrams are unambiguous. The diagrammatic proof does not depend on a particular expansion order, so it is expected to help the readers verify their evaluations of the effective potentials up to arbitrary orders.

Inspired by this proof, we have proposed a scheme to revise the super-daisy diagram resummation by deducting irrelevant terms at the connections of the ringlets to fit the Nielsen identity while averting the summation over all possible diagrams. However, a feasible algorithm is beyond the scope of this paper, and requires our further study.

\begin{acknowledgements}
We thank to Junmou Chen, Pyungwon Ko, Zhao-huan Yu, Hong-hao Zhang, Chengfeng Cai, Ligong Bian,  Chen Zhang, Gao-Liang Zhou, Ye-Ling Zhou,  for helpful discussions and communications. This work is supported in part by the National Natural Science Foundation of China under Grants No.12005312,  and the Sun Yat-Sen University Science Foundation. 

\end{acknowledgements}

\appendix

\newpage
\bibliography{DiagramNielsen}
\end{document}